\documentclass[preprintnumbers,superscriptaddress,amsmath,amssymb,nofootinbib,twocolumn,showpacs]{revtex4-1}
\usepackage{mathtools}
\usepackage{graphicx}
\usepackage{color}
\usepackage{natbib}
\usepackage{multirow}
\usepackage{SIunits}

\newcommand{\beq}{\begin{equation}}
\newcommand{\eeq}{\end{equation}}
\newcommand{\ben}{\begin{eqnarray}}
\newcommand{\een}{\end{eqnarray}}
\newcommand{\besub}{\begin{subequations}}
\newcommand{\eesub}{\end{subequations}}
\newcommand{\bi}{\begin{itemize}}
\newcommand{\ei}{\end{itemize}}
\newcommand{\nn}{\nonumber}
\newcommand{\bea}{\begin{align}}
\newcommand{\eea}{\end{align}}

\newcommand{\eg}{\mbox{\it e.g.}}

\newcommand{\pd}[2]{\frac{\partial#1}{\partial#2}}

\newcommand{\citeeq}[1]{Eq.~(\ref{#1})}

\newcommand{\citesec}[1]{Sect.~\ref{#1}}

\newcommand{\citeapp}[1]{App.~\ref{#1}}

\newcommand{\citetab}[1]{Tab.~\ref{#1}}
\newcommand{\citefig}[1]{Fig.~\ref{#1}}

\newcommand{\BIG}{{\sf BIG}}
\newcommand{\SLIM}{{\sf SLIM}}
\newcommand{\QUAINT}{{\sf QUAINT}}
\newcommand{\galprop}{\texttt{Galprop}}
\newcommand{\usine}{{\sc usine~v3.5}}
\newcommand{\minuit}{{\sc minuit}}
\newcommand{\minos}{{\sc minos}}
\newcommand{\usinebis}{{\sc usine}}

\def\aap{A\&A}%
\def\apj{ApJ}%
\def\apjl{ApJ}%
\def\apss{Ap\&SS}%
\def\jcap{J. Cosmology Astropart. Phys.}%
\def\jgr{J.~Geophys.~Res.}%
\def\mnras{MNRAS}%
\def\nat{Nature}%
\def\physrep{Phys.~Rep.}%
\def\prc{Phys.~Rev.~C}%
\def\prd{Phys.~Rev.~D}%
\def\prl{Phys.~Rev.~Lett.}%

\definecolor{myred}{RGB}{102,0,0}

\usepackage[pdftex,
            breaklinks=true,%
            colorlinks=true,%
            linkcolor=myred,
            urlcolor=blue,
            citecolor=blue,
            pdfauthor={Genolini et al.},%
            pdftitle={Cosmic-ray transport from AMS-02 B/C data: benchmark models and interpretation}%
           ]{hyperref}

\graphicspath{{./}{Figs/}}

\begin{document}

\title{Cosmic-ray transport from AMS-02 B/C data: benchmark models and interpretation} 
\author{Y. G\'enolini}
\email{yoann.genolini@ulb.ac.be}
\affiliation{Service de Physique Th\'eorique, Universit\'e Libre de Bruxelles,
  Boulevard du Triomphe, CP225, 1050 Brussels, Belgium}
\author{M. Boudaud}
\email{boudaud@lpthe.jussieu.fr}
\affiliation{LPTHE, Sorbonne Universit\'e \& CNRS, 4 Place Jussieu, 75252 Paris Cedex 05, France}
\author{P.-I. Batista}
\affiliation{Instituto de F\'isica de S\~ao Carlos, Universidade de S{\~a}o Paulo, CP 369,
  13560-970,  S{\~a}o Carlos, SP, Brazil}
\author{S. Caroff}
\affiliation{Sorbonne Universit\'es, UPMC Universit\'e Paris 06, Universit\'e Paris Diderot, Sorbonne Paris Cit\'e, CNRS, Laboratoire de Physique Nucl\'eaire et de Hautes Energies (LPNHE), 4 place Jussieu, F-75252, Paris Cedex 5, France}
\author{L. Derome}
\affiliation{LPSC, Universit\'e Grenoble Alpes, CNRS/IN2P3, 53 avenue des Martyrs, 38026 Grenoble,
  France}
\author{J. Lavalle}
\email{lavalle@in2p3.fr}
\affiliation{LUPM, CNRS \& Universit\'e de Montpellier (UMR-5299), Place Eug\`ene Bataillon,
  F-34095 Montpellier Cedex 05, France}
\author{A. Marcowith}
\affiliation{LUPM, CNRS \& Universit\'e de Montpellier (UMR-5299), Place Eug\`ene Bataillon,
  F-34095 Montpellier Cedex 05, France}
\author{D. Maurin}
\affiliation{LPSC, Universit\'e Grenoble Alpes, CNRS/IN2P3, 53 avenue des Martyrs, 38026 Grenoble,
  France}
\author{V. Poireau}
\affiliation{Univ. Grenoble Alpes, Univ. Savoie Mont Blanc, CNRS, LAPP, F-74940 Annecy, France}
\author{V. Poulin}
\affiliation{LUPM, CNRS \& Universit\'e de Montpellier (UMR-5299), Place Eug\`ene Bataillon,
  F-34095 Montpellier Cedex 05, France}
\author{S. Rosier}
\affiliation{Univ. Grenoble Alpes, Univ. Savoie Mont Blanc, CNRS, LAPP, F-74940 Annecy, France}
\author{P. Salati}
\affiliation{Univ. Grenoble Alpes, Univ. Savoie Mont Blanc, CNRS, LAPTh, F-74940 Annecy, France}
\author{P. D. Serpico}
\email{serpico@lapth.cnrs.fr}
\affiliation{Univ. Grenoble Alpes, Univ. Savoie Mont Blanc, CNRS, LAPTh, F-74940 Annecy, France}
\author{M. Vecchi}
\affiliation{Instituto de F\'isica de S\~ao Carlos, Universidade de S{\~a}o Paulo, CP 369,
  13560-970,  S{\~a}o Carlos, SP, Brazil}
\affiliation{KVI - Center for Advanced Radiation Technology, University of Groningen,
The Netherlands}

\date{\today}

\preprint{ULB-TH/19-03; LAPTH-020/19; LUPM:19-046}

\begin{abstract} 
  This article aims at establishing new benchmark scenarios for Galactic cosmic-ray propagation
  in the GV-TV rigidity range, based on fits to the AMS-02 B/C data with the \usine\
  propagation code. We employ a new fitting procedure, cautiously taking into account data systematic
  error correlations in different rigidity bins and considering Solar modulation potential
  and leading nuclear cross section as nuisance parameters.
  We delineate specific low, intermediate, and high-rigidity ranges that can be related to both
  features in the data and peculiar microphysics mechanisms resulting in spectral breaks.
  We single out a scenario which yields excellent fits to the data and includes all the presumably
  relevant complexity, the \BIG{} model. This model has two limiting regimes: (i) the
  \SLIM\ model, a minimal diffusion-only setup, and (ii) the \QUAINT\ model,
  a convection-reacceleration model where transport is tuned by non-relativistic effects. All
  models lead to robust predictions in the high-energy regime ($\gtrsim10$~GV), i.e. independent
  of the propagation scenario: at $1\sigma$, the diffusion slope $\delta$ is $[0.43-0.53]$, whereas
  $K_{10}$, the diffusion coefficient at 10~GV, is $[0.26-0.36]$~kpc$^2$~Myr$^{-1}$; we confirm
  the robustness of the high-energy break, with a typical value $\Delta_h\sim 0.2$. We also find a hint
  for a similar (reversed) feature at low rigidity around the B/C peak ($\sim 4$~GV) which might
  be related to some effective damping scale in the magnetic turbulence.
\end{abstract}

\maketitle
\tableofcontents

\section{Introduction}
\label{sec:intro}
The last decade in direct cosmic-ray (CR) detection experiments has been characterized by a
major improvement in the precision of the data available, and by an extension of the
covered dynamical range
\cite{AhnEtAl2009,PanovEtAl2009,AdrianiEtAl2014a,Ting2013,AdrianiEtAl2017,AmbrosiEtAl2017}.
In particular, with the AMS-02 data the community has to deal for the first time with
percent level precision and a welcomed redundancy in the measurements.

But, as well known, {\it great responsibility inseparably follows from great power}~\footnote{Or,
  in the original 1793 French version: ``{\it Une grande responsabilit\'e est la suite ins\'eparable
  d'un grand pouvoir}''.   French Revolution Parliamentary Archives, ``Tome 64 : Du 2 au 16 mai
  1793, S\'eance du mardi 7 mai 1793, page 287, available e.g. at
  https://frda.stanford.edu/fr/catalog/wx067jz0783\_00\_0293.}: since theoretical predictions are
very far from attaining that level of precision, both due to ignorance of the detailed underlying
microphysics (CR acceleration and transport) and because of irreducible limitations (e.g. due to
the intrinsic stochasticity of the sources \cite{Mertsch2011,GenoliniEtAl2017a}),
a preliminary question that should be addressed is that of the best strategy to take advantage
of such a wealth of data. In this paper, we primarily focus on the AMS-02 B/C data
\cite{AguilarEtAl2016a} and investigate how much they can constrain CR transport, aiming at
defining new benchmark models.

An ambitious approach would be to proceed with global fits of all available data,
attempting an overall and simultaneous understanding of CR sources (all species) and propagation.
However, this approach is prone to mixing uncertainties of different nature, with the risk
of devaluing the actual strength of the data by introducing poorly controlled parameters (for an
illustration, see \cite{CosteEtAl2012}). Since our current understanding of CR measurements has
more firm elements in the propagation part than in the source one, factorizing out propagation
effects from source effects, while inspecting their physical plausibility a posteriori, seems
justified. In this article, we proceed through partial tests of key aspects of the current
propagation paradigm, with the goal of validating it or highlighting its breakdown. An important
and relatively new issue in this area is that systematic errors are often dominant over statistical
ones. This requires a change of perspective in well-established practices of analyzing the data, as
well as new standards of rigor. It calls for establishing a satisfactory protocol for analyzing the
data on a relatively simple and homogeneous data sample. 

This article represents an important pillar in our CR data analysis based on the overall philosophy sketched above. As long established \cite{Owens1976a,GinzburgEtAl1980,ProtheroeEtAl1981,StrongEtAl1998,MaurinEtAl2001}, a flux ratio of elements present but in traces in the solar system material and interstellar medium (ISM), such as Lithium, Beryllium, Boron (``secondaries''), to abundant species like Carbon or Oxygen (``primaries'') is extremely sensitive to propagation parameters. It was shown to be also almost insensitive to the energy spectrum of the injected primary species, notably if those are described by a common power-law in rigidity \cite{MaurinEtAl2002a,PutzeEtAl2011,GenoliniEtAl2015}, a rather generic prediction of studies of CR acceleration at sources \cite{MalkovEtAl2001,CaprioliEtAl2014,Amato2014a,MarcowithEtAl2016}. In particular, at high rigidities we expect the B/C ratio (currently the most precisely measured) to be dominantly affected by diffusive propagation and nuclear cross sections.

In~\cite{GenoliniEtAl2017}, we performed an analysis of the high-rigidity range of the AMS-02 B/C ratio \cite{AguilarEtAl2016a}, finding evidence for a diffusive origin of the observed spectral break, at the same rigidity scale inferred from a similar feature in the proton and helium CR fluxes \cite{AguilarEtAl2015}, i.e. $\sim 300$~GV. Actually, recent years have been characterized by the observational establishment of ``spectral anomalies'' (for reviews, see~\cite{Serpico2015,Serpico2018}), in particular of spectral breaks in primary species \cite{PanovEtAl2009,AhnEtAl2010,AdrianiEtAl2011a,AguilarEtAl2015,AguilarEtAl2015a,AguilarEtAl2017}. In turn, there has been growing evidence in favor of their interpretation in terms of a high-rigidity break in the diffusion coefficient~\citep{GenoliniEtAl2017,ReinertEtAl2018,XueEtAl2019}, notably after the first AMS-02 publications of nuclear CR fluxes~\cite{AguilarEtAl2017,AguilarEtAl2018}. In this article, we move several steps beyond our previous analysis \cite{GenoliniEtAl2017}, presenting a {\it complete} analysis aiming at constraining CR propagation and at proposing new benchmark setups: First, we rely on an improved analysis of the B/C data by the AMS-02 collaboration \cite{AguilarEtAl2018}. We further benefit from additional data on the primary species to constrain the break independently from the B/C ratio---using the C and O fluxes \cite{AguilarEtAl2017} which are most contributing species to B production \cite{GenoliniEtAl2018}, but were not available to Ref.~\cite{GenoliniEtAl2017}. Second, we follow the new methodology proposed in \cite{DeromeEtAl2019} to analyze the AMS-02 data, carefully accounting for a number of subtle (but highly important) effects which are usually ignored, notably the (partial) correlations in systematic errors. This approach enables a straightforward and sound statistical interpretation of the models (best-fit models have $\chi^2/{\rm dof}\sim 1$), also allowing for their inter-comparison. Third, we propose a new generic propagation model (dubbed \BIG\ in the following), with a number of parameters that should be sufficient to describe all key features currently present in the data. In addition to a high-rigidity break, a modification of the diffusion coefficient at low rigidity is enabled ($\lesssim 5$ GV), with two limiting cases (dubbed \SLIM\ and \QUAINT): this allows us to assess the relative discriminating strength of the data in this energy range and to shed new light on propagation in the low-rigidity regime, where a second diffusion break might be present.

The paper is organized as follows: In \citesec{sec:def_bench}, the one-dimensional (1D) propagation model and the essential physical effects involved in CR propagation are presented, before introducing our three benchmark scenarios (\BIG, \SLIM, and \QUAINT). In \citesec{sec:fit_bc}, we describe the specific iterative procedure used for the B/C analysis, and this procedure is validated and checked in two appendices: \citeapp{app:consistency} assesses the robustness of the derivation of the high-energy break, by taking advantage or not of the C and O fluxes; \citeapp{app:low_rig} further discusses the dependence of the fit parameters upon the lower rigidity cut, to better illustrate and give meaning to the terms ``low-rigidity'' and ``high-rigidity''parameters. Best-fit results for our three scenarios are presented in \citesec{sec:results}, where these scenarios are also tentatively interpreted in terms of the underlying microphysics. In \citesec{sec:conclusion}, we report our conclusions and mention natural follow-up works. Note that all results are obtained for a 1D model of our Galaxy, with the size of the diffusive halo fixed. In order to allow for a broader usage of our results, as they may have some consequence in predicting the fluxes of other secondary species, \citeapp{app:Ldiff} provides a scaling of the high-rigidity parameters with $L$, whereas \citeapp{app:dict} reminds the reader of the effectiveness of this description, and provides a ``dictionary'' to interpret the results in terms of a two-dimensional (2D) model with different halo sizes.
\section{Transport models}
\label{sec:def_bench}
In this section, we introduce the generic propagation equation that we further solve
semi-analytically in the framework of the \usinebis\ code \cite{Maurin2018} -- for fully numerical
frameworks complementary to ours, we refer the reader
to Refs. \cite{StrongEtAl1998,EvoliEtAl2008,EvoliEtAl2017,Kissmann2014,KissmannEtAl2015}.
We also set a generic CR transport configuration motivated by theoretical arguments
on the microphysics of CRs, with a focus on possibly important low-energy processes.
This generic setup will itself be used as a benchmark configuration, of which we shall explore
two limiting regimes. These three cases characterize new benchmark models (dubbed \BIG, \SLIM,
and \QUAINT---see \citesec{ssec:benchmarks}) that are aimed at capturing different theoretical
assumptions, while still being data driven. These configurations will be shown to provide excellent
fits to the current B/C data assuming simple power-law primary CR spectra. Not only may this
stimulate further microphysical interpretations, but it also offers a basis for a description of
other CR data, like for instance the positron
\cite{DelahayeEtAl2009,DelahayeEtAl2010,BoudaudEtAl2017a} and antiproton fluxes
\cite{DonatoEtAl2001,GiesenEtAl2015,ReinertEtAl2018,BoudaudEtAl2019}. Eventually, these benchmarks
will be instrumental in characterizing and hopefully reducing theoretical uncertainties entering
searches for exotic phenomena
(\eg~\cite{DelahayeEtAl2008,BoudaudEtAl2015a,BoudaudEtAl2015,BoudaudEtAl2017,BoudaudEtAl2018,BoudaudEtAl2018b,ReinertEtAl2018}).

The {\em min}, {\em med}, and {\em max} benchmark values proposed in Ref.~\cite{DonatoEtAl2004},
all based on the same model, were defined to roughly bracket the theoretical uncertainties on
dark matter-induced antiproton flux predictions assuming the best-fitting propagation parameters
of Ref.~\cite{MaurinEtAl2001}. However, these values have been challenged by a series of
complementary constraints
\cite{LavalleEtAl2014,BoudaudEtAl2017a,ReinertEtAl2018}, and are anyway no longer consistent with
the B/C data~\cite{GenoliniEtAl2017}. The revised reference models we propose here rely on
different assumptions on the microphysics of CR transport instead of different values of
parameters within the same configuration. This change in philosophy stems from the fact that
with increasing precision in the (multiwavelength and multimessenger) observational data and
improvements on the theory side, we expect to arrive soon to a much better understanding and
description of the CR microphysics itself than it was possible two decades
ago~\cite{MarcowithEtAl2016,AmatoEtAl2018}.
\subsection{Transport description}
\label{ssec:transport}
\subsubsection{Transport equation}
\label{sssec:prop_eq}
The general formalism that provides a powerful description of the transport of CRs in the
Milky Way derives from the seminal textbook by Ginzburg and Syrovatskii \cite{GinzburgEtAl1964}
(see also \cite{BerezinskiiEtAl1990,Schlickeiser2002}), and relies on the following
diffusion-advection equation for a CR species of index $\alpha$, here in the steady-state
approximation and in energy space (rather than rigidity or momentum space):
\begin{widetext}
\ben
\label{eq:prop}
& - \;\vec{\nabla}_{\bf x} \left\{ K(E)\,\vec{\nabla}_{\bf x}\psi_\alpha -
\vec{V}_{\rm c} \psi_\alpha \right \} 
+ \pd{}{E} \left\{ b_{\rm tot}(E)\;\psi_\alpha - \beta^2\, K_{pp}\,\pd{\psi_\alpha}{E} \right\}
+ \sigma_{\alpha}\,v_\alpha\, n_{\rm ism}\, \psi_\alpha + \Gamma_{\alpha}\,\psi_\alpha\nn\\
& = \;q_{\alpha}+\sum_{\beta} \left\{ \sigma_{\beta\to \alpha}v_\beta n_{\rm ism}\,+\Gamma_{\beta\to\alpha}
\right\}\,\psi_\beta \;.
\een
\end{widetext}
This equation describes the spatial and energy evolution of the differential interstellar CR
density per unit energy $\psi_\alpha\equiv dn_\alpha/dE$, assuming a net primary injection rate of
$q_\alpha$, and a secondary injection rate arising from inelastic processes converting heavier
species of index $\beta$ into $\alpha$ species (with a production rate $ 
\sigma_{\beta\to \alpha}v_\beta n_{\rm ism}$ on the ISM density $n_{\rm ism}$, or a decay rate
$\Gamma_{\beta\to\alpha}$). This source term is balanced by several other terms, among which the decay
rate $\Gamma_{\alpha}$ (if relevant). The central piece of the propagation equation is the spatial
diffusion coefficient $K$, that we discuss in more detail in \citesec{sssec:kdiff}. The other
processes are mostly relevant at low rigidity, but may still affect the determination of
higher-energy parameters: convection is featured by a velocity $\vec{V}_{\rm c}$, diffusive
reacceleration is parameterized by the energy-dependent coefficient $K_{pp}$, and the inelastic
destruction rate is given by $\sigma_{\alpha}v_\alpha n_{\rm ism}\, \psi_\alpha$, with the  $\sigma$'s
being energy-dependent nuclear cross sections; energy losses are characterized by the rate
$b_{\rm tot}\equiv dE/dt$, which includes ionization and Coulomb processes as prescribed
in~\cite{MannheimEtAl1994,StrongEtAl1998}, as well as adiabatic losses induced by convection and
reacceleration, see~\cite{MaurinEtAl2002a,PutzeEtAl2010}.

Finally, we can switch from the interstellar (IS) CR flux predictions to the top-of-atmosphere (TOA)
ones by means of the force-field approximation \cite{GleesonEtAl1968a,Fisk1971}, for which we only
indicate the Fisk potential $\phi_{\rm F}$. The latter is constrained from Ref.~\cite{GhelfiEtAl2016}
for the AMS-02 data taking period.
\subsubsection{Geometry and cross sections}
\label{sssec:geom}
We assume a 1D propagation model, as introduced in
\eg~\cite{BulanovEtAl1974,Ptuskin1974,Jones1979,PtuskinEtAl1990,JonesEtAl2001,MaurinEtAl2001},
where the magnetic halo confining the CRs is an infinite slab in the radial direction and of
half-height $L$. Indeed, the radial boundary has only a minor quantitative impact on other transport
parameters when the diffusion coefficient is taken as a scalar function (see
\eg~\cite{MaurinEtAl2002a,PutzeEtAl2010}), and neglecting the radial dependence allows us to more
efficiently probe the entire available parameter space without significant loss of generality.
Therefore, we consider the vertical coordinate $z$ to be the only relevant spatial coordinate. See
however \citeapp{app:dict} for some considerations on the correspondence between 1D and 2D models.

The sources of CRs and the ISM gas which they scatter off are taken homogeneous in an infinitely
thin disk at $z=0$, with an effective half-height $h=100$ pc. Energy losses are also considered to
be localized in the disk, $b_{\rm tot}\propto 2\,h\,\delta(z)$. The ISM density is set to
$n_{\rm ism}=2\,h\delta(z)\,n_0$, where $n_0=1$ cm$^{-3}$, corresponding to a surface density of
$\Sigma_{\rm ism}= 2\,h\,n_0\simeq 6\times 10^{20}$ cm$^{-2}$ consistent with observations
\cite{Ferriere2001}. We assign 0.9 and 0.1 of this budget to hydrogen and helium (in number),
respectively. We do not indulge here in a more detailed discussion of the determination of these
parameters and of their error from independent observations, since they are largely degenerate with
the normalization of the diffusion coefficient (see, {\em e.g.}, \cite{MaurinEtAl2010}).

For the nuclear production and spallation cross sections, we use as reference the sets of tables
from the \href{https://galprop.stanford.edu/}{\galprop} package, while allowing for normalization,
energy shift and low-energy slope changes according to the NSS method described in
\cite{DeromeEtAl2019}.
\subsubsection{Diffusion in real space and momentum space, and convection}
\label{sssec:kdiff}
An important physical ingredient to all propagation models is the diffusion coefficient, which
describes the scattering of CRs off magnetic turbulence. We assume that it can be taken as a
scalar function, homogeneous and isotropic all over the magnetic slab. This is likely a good
approximation in the context of B/C analyses because the CR flux is locally isotropic and
the magnetic field configuration of the Milky Way exhibits relative fluctuations $\delta B/B\sim 1$
over all relevant spatial scales $\lambda$ (or wavenumbers $k=2\pi/\lambda$)
\cite{CasseEtAl2002,SunEtAl2008a,JanssonEtAl2012a}. Since at the energies
of interest, the CR flux is expected to be contributed to by many sources mostly located many
``magnetic domains'' away, an isotropic diffusion should provide at least a reasonable effective
description of the data \cite{BerezinskiiEtAl1990}. For discussions on anisotropic models, see
\eg~Refs.~\cite{DeMarcoEtAl2007,EvoliEtAl2012a}.

On the theory side, it is expected that the magnetic turbulence responsible for CR diffusion has
different scaling behaviors in $k$-space, as a consequence of various possible phenomena. For
instance, the turbulence power spectrum can be dominated by different sources depending on the
dynamical range, with the resulting ``two-zone'' models known to provide good fits to the
data~\cite{Tomassetti2012,GuoEtAl2018} (see also \cite{SeoEtAl1994,JonesEtAl2001}). A very
appealing scenario is proposed in
Refs.~\cite{BlasiEtAl2012a,AloisioEtAl2013,AloisioEtAl2015,EvoliEtAl2018}, relying on streaming
instability \cite{Wentzel1969,Holmes1975,Skilling1975}, where at rigidities beyond a few hundreds
of GV, CRs diffuse on the turbulence injected on large scales by supernova bubbles. This turbulence
cascades down until crossing the rigidity scale where the turbulence induced by CRs themselves
takes over. This naturally generates a break like the one observed in the CR spectra and discussed
just above. On the other hand, it is known that the CR spectra observed at low rigidity by the
Voyager I spacecraft~\cite{CummingsEtAl2016,WebberEtAl2017} have a spectral slope rather different
from the slope at intermediate rigidities. Due to the CR-wave coupling, any phenomenon with a
low-rigidity characteristic scale, affecting either propagation or injection, may thus be at the
origin of correlated changes in the CR spectra and the diffusion coefficient. For propagation, such
a scale might arise due to the decrease of the CR pressure as CRs get closer and closer to the
nonrelativisitic regime, and/or be related to some dissipation of the turbulence power spectrum
\cite{YanEtAl2004,PtuskinEtAl2005,PtuskinEtAl2006a,ShalchiEtAl2010,EvoliEtAl2014,XuEtAl2016}.
In the following, while remaining agnostic on these specifics, and in contrast to previous B/C
studies performed in the context of semi-analytical models (\eg~\cite{MaurinEtAl2001,MaurinEtAl2002a,PutzeEtAl2010,PutzeEtAl2011}), we want to capture
the possibility that the diffusion coefficient departs from a single power law. This is justified
by both theoretical arguments and observational evidence, as recalled above. 

Starting from general considerations arising in the quasi-linear theory
(\eg~\cite{Jokipii1966,BerezinskiiEtAl1990,Schlickeiser2002,Shalchi2009}), the diffusion
coefficient is expected to be linked to the magnetic turbulence spectrum $|\delta B/B|$ through
\ben
\label{eq:k_linear_th}
K(E) = \frac{v\,l_{\rm mfp}}{3} \simeq \frac{v}{3}\frac{r_{\rm L}}{|\delta B/B|_{k_{\rm L}}^2}\,,
\een
where $v$ is the CR speed, $l_{\rm mfp}$ is the mean free path length, $r_{\rm L}\propto R/B$ is
the Larmor radius defined from the rigidity $R=p/Z$, and $k_{\rm L}\propto 1/r_{\rm L}$ is the
turbulence mode in resonance with the CR Larmor radius. Consequently, we propose a general form
for the diffusion coefficient that can account for breaks in both the high-rigidity range and the
low-rigidity range (hidden in the factor $r_{\rm L}/|\delta B/B|_{k_{\rm L}}^2$ above), which
reads
\begin{widetext}
\ben
\label{eq:def_K}
K(R) = \underbrace{\beta^\eta}_{\text{non-relativistic regime}} \, K_{10} \,
\underbrace{\left\{ 1 + \left( \frac{R}{R_{\rm l}} \right)^{\frac{\delta_{\rm l}-\delta}{s_{\rm l}}}
  \right\}^{s_{\rm l}}}_{\text{low-rigidity regime}}\,
\underbrace{\left\{  \frac{R}{\left(R_{\rm 10}\equiv 10\,{\rm GV}\right)} \right\}^\delta}_{\text{intermediate regime}} \,
\underbrace{\left\{  1 + \left( \frac{R}{R_{\rm h}} \right)^{\frac{\delta-\delta_{\rm h}}{s_{\rm h}}}
  \right\}^{-s_{\rm h}}}
_{\text{high-rigidity regime}}\,.
\een
\end{widetext}
In the above equation, $\beta=v/c$ is the dimensionless CR speed, and $R_{\rm l/h}$ is the location
of the low/high-rigidity break, while $R_{\rm 10}$ is an intermediate rigidity (here taken at 10 GV
on purpose) such that $R_{\rm l}<R_{10}<R_{\rm h}$ ($R_{\rm l}\ll R_{\rm h}$). We then get the scaling
$K(R)\propto \beta^\eta R^{\delta_{\rm l}}$ in the limit $R\ll R_{\rm l}$, and the scaling
$K(R)\propto R^{\delta_{\rm h}}$ in the limit $R\gg R_{\rm h}$. Therefore, $\delta_{\rm l}$, $\delta$, and
$\delta_{\rm h}$ simply describe the diffusion spectral indices in the low-, intermediate-, and
high-rigidity regime, respectively. The parameter $s_{\rm l}$ ($s_{\rm h}$) characterizes how fast
the spectral change proceeds around $R_{\rm l}$ ($R_{\rm h}$), and is inspired by the need to
describe the very smooth hardening of the B/C data showing up at high rigidity. Indeed, we recall
that the previous high-rigidity analysis performed in Ref.~\cite{GenoliniEtAl2017} provided 
support to a softening of the diffusion coefficient to explain this feature, such that we can
already anticipate that $\delta_{\rm h}<\delta$.
The normalization of the diffusion coefficient $K_{10}$ (which carries the physical units) is
another free parameter. Mind the difference with the convention used in most past
analyses, where the normalization was instead $K_0$ and was defined at a rigidity
$R_0=1\,{\rm GV}$. Note also that $K_{10}\simeq K(10\,{\rm GV})$, not a strict equality, because
of the influence of the other terms.

We further introduce the spectral-change parameters
\besub
\label{eq:Delta}
\ben
\Delta_{\rm l} &=& \delta - \delta_{\rm l} \,, \\
\Delta_{\rm h} &=& \delta -  \delta_{\rm h}\,,
\een
\eesub
As already mentioned above, $\Delta_{\rm h}$ is expected to be positive. A positive $\Delta_{\rm l}$
is also expected from damping arguments and from the flattening of the primary CR spectra observed
by Voyager I, as CRs may diffuse mostly on self-generated turbulence---see the discussion above.
Notice that in the low-rigidity regime, additional non-relativistic processes might further
be considered in an effective way by raising the velocity $\beta$ to the power $\eta$, an effective
index which---it has been argued---might take negative values in some regimes
\cite{PtuskinEtAl2005,PtuskinEtAl2006a}. 

A comment on $\eta$ is in order: since the rigidity range of CR data analyzed in this article is
always relativistic, sizable departures from $\eta=1$ (the natural value from quasi-linear
theory, see \citeeq{eq:k_linear_th}) and/or large values of $V_{\rm A}$ (which allows for energy
redistribution) may be needed to affect appreciably CRs whose $v\simeq c$. In a certain sense, 
$\eta$ is thus not a very valuable {\em effective} parameter for the problem at hand. Nonetheless,
we keep the $\eta$ parameter in the discussion for historical reasons, since in combination with
strong reacceleration it used to be an important ingredient in past studies of B/C data, notably at
low energies~\cite{SeoEtAl1994,StrongEtAl1998,JonesEtAl2001,MaurinEtAl2001,DonatoEtAl2004}.
Sufficiently large negative values of $\eta<-\delta_{\rm l}$ (or, similarly, of $\delta_{\rm l}<-1$ if
$\eta=1$) can also imply superluminal diffusion \cite{DunkelEtAl2007,AloisioEtAl2009} in the
non-relativistic regime. In this sense, we caution the reader never to extrapolate a-critically
the functional forms obtained here too far from the rigidity range over which the fits have been
obtained.

Let us now be more specific about reacceleration. It turns out that spatial diffusion can rather
generically be linked to diffusion in momentum space (aka reacceleration) in most (but not all)
cases \cite{BerezinskiiEtAl1990}. We include diffusion in momentum space through an additional
diffusion coefficient $K_{pp}$---see \citeeq{eq:prop}. We follow the reacceleration model proposed
in Refs.~\cite{OsborneEtAl1988,SeoEtAl1994,JonesEtAl2001}, which is implemented in \usinebis\ such
that $K_{pp}(R,\vec{x})=2\,h\,\delta(z)\,K_{pp}(R)$, and
\ben
\label{eq:Kpp}
K(R)\times K_{pp}(R) = \frac{4}{3}\,V_{\rm A}^2 \,\frac{p^2}{\delta\,(4-\delta^2)(4-\delta)}\,,
\een
where $V_{\rm A}$ is an effective Alfv\'enic speed characterizing the magnetic turbulence---$\delta$
is the diffusion spectral index in the intermediate inertial regime. Since it appears explicitly
only as a normalization factor, we stick to this formula even when spatial diffusion exhibits
several spectral regimes. The fact that reacceleration is {\em effectively} localized in the disk
allows us to partly solve \citeeq{eq:prop} analytically, which significantly speeds up the numerical
exploration of the parameter space \cite{Maurin2018}. While this ``pinching'' is a fair approximation
for ionization and Coulomb processes, it is only a convenient approximation for adiabatic losses
induced by convection and reacceleration. Hence, care should be taken when comparing inferred values
of the parameters $V_{\rm c} \equiv | \vec{V}_{\rm c}|$ and $V_{\rm A}$ with theoretical expectations.
Loosely speaking, one can expect to recover
the phenomenology of a more extended reacceleration zone by a rescaling of $V_{\rm A}^2$ by a factor
$h/z_{\rm A}$ \cite{MaurinEtAl2002}, where $z_{\rm A}$ is the half-height over which reacceleration would spread in the
magnetic slab \cite{JonesEtAl2001}. So, for $h/z_{\rm A}\simeq {\cal O}(h/L)$, our fitted value of
$V_{\rm A}$ should be scaled by a factor $\sqrt{L/h}$ before any comparison against theoretical or observational
constraints~\cite{ThornburyEtAl2014,DruryEtAl2017}.

Finally, convection also arises quite naturally in the framework discussed above. We include
convection in the standard way by means of the convection velocity
\ben
\label{eq:vc}
\vec{V}_{\rm c}(z) = \frac{z}{|z|}\,V_{\rm c}\,\vec{e}_z\,,
\een
where $z$ is the vertical coordinate and $\vec{e}_z$ the unit vector along the vertical axis
crossing the magnetic slab of extension $[-L,L]$ along that axis.

\subsection{Benchmark models}
\label{ssec:benchmarks}
In the most general case,  the free parameters featuring the propagation modeling that we have
introduced above are the following: $L$ for the magnetic halo size; $K_{10}$, $\delta$, $\eta$,
$R_{\rm l}$, $\delta_{\rm l}$ (equivalently $\Delta_{\rm l}$), $s_{\rm l}$, $R_{\rm h}$, $\Delta_{\rm h}$
(equivalently $\delta_{\rm h}$), and $s_{\rm h}$  for the diffusion coefficient; $V_{\rm A}$ for
reacceleration; $V_{\rm c}$ for convection.
This is a 12-parameter space, hence a huge configuration volume to explore.

Based on previous studies, we can further fix $L$ which is highly correlated with $K_{10}$, see discussion in \citeapp{app:Ldiff}. Unless specified otherwise, we will set $L$ to 10
kpc in the following. Moreover, as anticipated in \citesec{sec:intro}, the determination of (an
interval for) the three parameters describing the high-rigidity break benefits from fits including
primary species, see \citesec{sec:fit_bc}. Finally, without loss of generality, we fix the
smoothing low-rigidity break parameter $s_{\rm l}=0.05$, which amounts to consider a fast transition.
This is however not critical to the fit. Hence, we are left with 7 free parameters.

From these 7 parameters, we design three different benchmark propagation models which may be related
to quite different limiting regimes of the underlying microphysics. The first, most generic,
model includes the whole setup introduced above: let us name it the \BIG\ model.
The second one is much simpler as it is free of convection and reacceleration, hence with much
less free parameters, while providing fits to the data comparable to the previous one
(see \citesec{sec:results}); let us call it the \SLIM\ model. The third and last one
includes both reacceleration and convection, but relates the possible change in the propagation at
low rigidities to a change originating specifically in the non-relativistic regime ($\eta$),
instead of a more generic low-rigidity break in the diffusion coefficient. This scenario provides a
slightly worse fit to the data compared to the previous ones, at the expense of a large
reacceleration $V_{\rm A}$. However, it allows us to connect the current analysis to the strong
reacceleration models that were popular in the past; let us dub it the \QUAINT\ model. Both the
\SLIM\ and \QUAINT\ models are actually particular cases of the \BIG\ model, but put the emphasis on
different physical processes at low rigidity. In the following, we provide the details of these
three configurations.

\subsubsection{\textsf{BIG}: the paradigmatic model}
\label{sssec:big}
The \BIG\ model includes a double-break diffusion coefficient, as well as convection
and reacceleration. Its minimal version fixes the non-relativistic parameter $\eta=1$, while a
non-minimal configuration may allow $\eta$ to vary. The latter case will actually help justify
the former one independently from theoretical arguments. Therefore, the \BIG\ model stands
for the most general configuration describing the propagation equation, \citeeq{eq:prop},
and which allows us to probe the low-rigidity processes with the largest flexibility and
complexity. This model has a total of 6 (7)  parameters in the minimal (non-minimal) configuration,
which are recalled in \citetab{tab:free_params}.
\subsubsection{\textsf{SLIM}: the minimal (double-break diffusion) model}
\label{sssec:slim}
The \SLIM\ model is a subpart of \BIG, which discards convection and reacceleration as major
players at low rigidity ($V_{\rm A}=V_{\rm c}=0$ km/s), but instead insists on relating low-rigidity
features to changes in the magnetic turbulence properties. It also assumes a standard scaling
in the non-relativistic regime, with $\eta=1$. This model, though very minimal, will be shown to
provide an excellent fit to the data. It has 4 free parameters which are summarized in
\citetab{tab:free_params}. Note that an important advantage of this model is that it comes with a
fully analytical solution to the transport equation. This is particularly attractive in the context
of dark matter predictions \cite{MaurinEtAl2002,LavalleEtAl2012}.

\subsubsection{\textsf{QUAINT}: the `old-fashion' strong reacceleration model}
\label{sssec:quaint}
Our last benchmark model is the \QUAINT\ model, which is also a subpart of the \BIG\ model, and
which aims at describing the low-rigidity features mostly in terms of reacceleration and
convection. This model is actually the direct descendant of the
{\em min}-{\em med}-{\em max} models \cite{MaurinEtAl2001,DonatoEtAl2004} as it relies on almost
the same configuration space, except for the high-rigidity break in the diffusion coefficient
(which was not observed at the time of its ancestors and will anyway be treated as a nuisance
parameter in the statistical analysis). Large $V_{\rm A}$, in combination with a non-trivial value
of $\eta\lesssim 0$ is needed to provide decent fits to the data. A large $V_{\rm A}$ in turn
couples low-rigidity and high-rigidity features, maximizing parameter correlations.  The \QUAINT\
model has 5 free parameters, made explicit in \citetab{tab:free_params}. In practice, the
diffusion coefficient associated with the \QUAINT\ model is that of \citeeq{eq:def_K} without
the low-rigidity term.

\begin{table}
  \begin{center}
    \begin{tabular}{| c || c | c | c | }
      \hline\hline
  Free parameters / Models & {\bf \sf BIG} & {\bf \sf SLIM} & {\bf \sf QUAINT} \\
 \hline\hline
 $K_{10}$ & \checkmark & \checkmark &  \checkmark \\
 $\delta$ & \checkmark &  \checkmark  & \checkmark \\
\hline
 $\eta$ & 1 or \checkmark &  1 &  \checkmark \\
 $\delta_{\rm l}$ & \checkmark & \checkmark & N/A \\
 $s_{\rm l}$ & 0.05 &  0.05 & N/A \\
 $R_{\rm l}$ & \checkmark & \checkmark & N/A \\
\hline
$V_{\rm A}$ & \checkmark & N/A  & \checkmark \\
$V_{\rm c}$ & \checkmark & N/A  & \checkmark \\
 \hline \hline
    \end{tabular}
    \caption{Free parameters of the three benchmark models \BIG, \SLIM, and
      \QUAINT. The first block of parameters is associated with the diffusion coefficient
      in the intermediate regime, and is common to all models. The second block
      is related to a potential low-rigidity break in the diffusion coefficient, or to purely
      non-relativistic effects. The last block is related to reacceleration and convection.}
    \label{tab:free_params}
  \end{center}
\end{table}

\section{Fitting strategy}
\label{sec:fit_bc}
In this section, we explain the fitting strategy used to extract the benchmark propagation
parameters for the models presented above (\BIG, \SLIM, and \QUAINT).
Fits are performed with the \minuit\ package \cite{JamesEtAl1975} interfaced
with the \usinebis\ code \citep{Maurin2018}, and in particular, asymmetric error bars
on the parameters rely on the \minos\ algorithm. For more technical details and subtleties
on the setup and the analysis, we refer the reader to Ref.~\citep{DeromeEtAl2019}.
\subsubsection{Modeling uncertainties}
For each run, the fluxes of the elements from Beryllium (Be) to Silicon (Si) are computed assuming
that $^{10}$B, $^{11}$B (and $^{10}$Be, decaying into $^{10}$B) are pure secondary species and that
all the heavier elements contain a secondary and a primary component. We assume the primary
injection to follow a universal power law in rigidity with index $\alpha$. The secondary component
is computed by a full spallation network using the \galprop\ cross-section parameterization
(see appendices of \citep{GenoliniEtAl2018}). It has been shown in \citep{DeromeEtAl2019} that
this parameterization provides the best agreement with the data, and that uncertainties on
spallation cross sections are satisfactorily taken into account using only the
$^{12}{\rm C}+{\rm H}\rightarrow {}^{11}{\rm B}$ production cross section as nuisance parameter with
the ``normalization, slope and shape'' (NSS) strategy. For each
run, the {\it initial} default procedure is to fix the normalization of the primary components of
all elements to the 10.6 GeV/nuc data point of HEAO-3~\citep{Maurin2018}, except for the CNO
elements which affect more directly the B/C ratio: The latter ones are normalized to the
C, N, O data of AMS-02 at a rigidity of 50 GV. The power-law index $\alpha$ is first set to 2.3,
and fixed later via the iterative procedure explained below.  The solar modulation of CRs is
described in the force-field approximation, for which the Fisk potential $\phi_{\rm F}$
is averaged over the AMS-02 B/C data taking period. Based on \citep{GhelfiEtAl2016}, we set $\phi_{\rm F}$ as a nuisance parameter of mean value 730~GV and dispersion $\sigma_{\phi_{\rm F}}=$100~MV.

\subsubsection{Data errors}
The AMS-02 collaboration does not provide users with the covariance error
matrix of the data. In this case, it is common practice to estimate the \textit{total} errors by
summing systematics and statistics in quadrature. This procedure is however inappropriate when
systematics dominate, i.e. below $\sim$ 100 GV for the AMS-02 data, and for which correlations
in energy are expected to be important. A major novelty of the present analysis is to perform fits taking into account these correlations with a parametric form of the covariance matrix. The matrix was built thanks to the information provided in the Supplemental Material of the AMS-02 B/C analysis \cite{AguilarEtAl2018}. In particular, the different systematics, which are associated with different physics processes in the detector, have different correlation lengths, and the covariance matrix built reflects this complexity. For more precision we refer the reader to~\citep{DeromeEtAl2019}.

%
%
\subsubsection{C and O primaries}
It has been noted already that recent data show an indication for a
high-rigidity break in the diffusion coefficient. However, these data are at present still far
for providing us with the precise characteristics of this feature. In fact, in order to start
gaining statistical confidence in the very existence of this break, the typical strategy until
now has been to combine the B/C data with independent indications for the break. For instance,
in~\citep{GenoliniEtAl2017} we used the AMS-02 $p$ and He data to that purpose.

As anticipated in \citesec{sec:intro}, we do not any longer focus our B/C analysis on the
high-rigidity regime. Instead, we want to provide reference
values for the parameters controlling the low- and intermediate-rigidity regimes. Consequently,
it is a natural choice to use the high-rigidity break parameters as {\it nuisance} parameters.
However, in order to establish the plausible range over which to vary them, it is recommended to
resort to complementary and ``independent'' input. To minimize possible biases due for instance to
possibly different origins of the different species, we choose to limit ourselves to the C and O
fluxes because:
\bi
\item They are by far the main progenitors of the B and C fluxes entering the B/C ratio.
\item Fitting them allows us to determine a plausible value of the common spectral index of
  nuclei $\alpha$ as well as to check their {\it consistency} with the parameters obtained with
  the B/C analysis. Indeed, although we focus here on the B/C observable to determine the
  propagation parameters, we still want to make sure that our results are consistent with the
  observed primary fluxes.
\ei
In \citeapp{app:consistency}, we make the important sanity check of neglecting this external input,
relying solely on the B/C data to determine both the high-rigidity spectral break (and the
other propagation parameters). We show that the obtained results are perfectly consistent with the
``factorizing'' procedure sketched above, at the obvious price of a worse determination of the
propagation parameters. This is indeed not surprising, since the B flux, which dominates the B/C
statistical error at high rigidity, is more than one order of magnitude scarcer than the C and O
ones.
\subsubsection{The fitting procedure}
The technical implementation of the fits proceeds by iteration. After fixing the (low- and
intermediate-rigidity) propagation parameters with a first fit of the B/C ratio (as described
above, i.e. with $\phi_{\rm F}$ and the $^{12}{\rm C}+{\rm H}\rightarrow {}^{11}\rm B$ production
cross section taken as nuisance parameters), we perform a combined fit of the AMS-02 C and O
fluxes keeping the following parameters as free parameters: source-term normalizations,  
power-law dependence in rigidity $\alpha$, and break parameters ($R_{\rm h}$, $s_{\rm h}$,
and $\Delta_{\rm h}$). We then use the best-fit values of the break parameters
and associated covariance matrix as nuisance parameters in
a new B/C fit, keeping also $\alpha$ fixed to its best-fit value. In practice, only a couple of
iterations are needed to get the parameters compatible between two consecutive iterations. The
results discussed below are the outcome of this procedure.

\begin{figure}[!htb]
  \centering
  \includegraphics[width=\columnwidth]{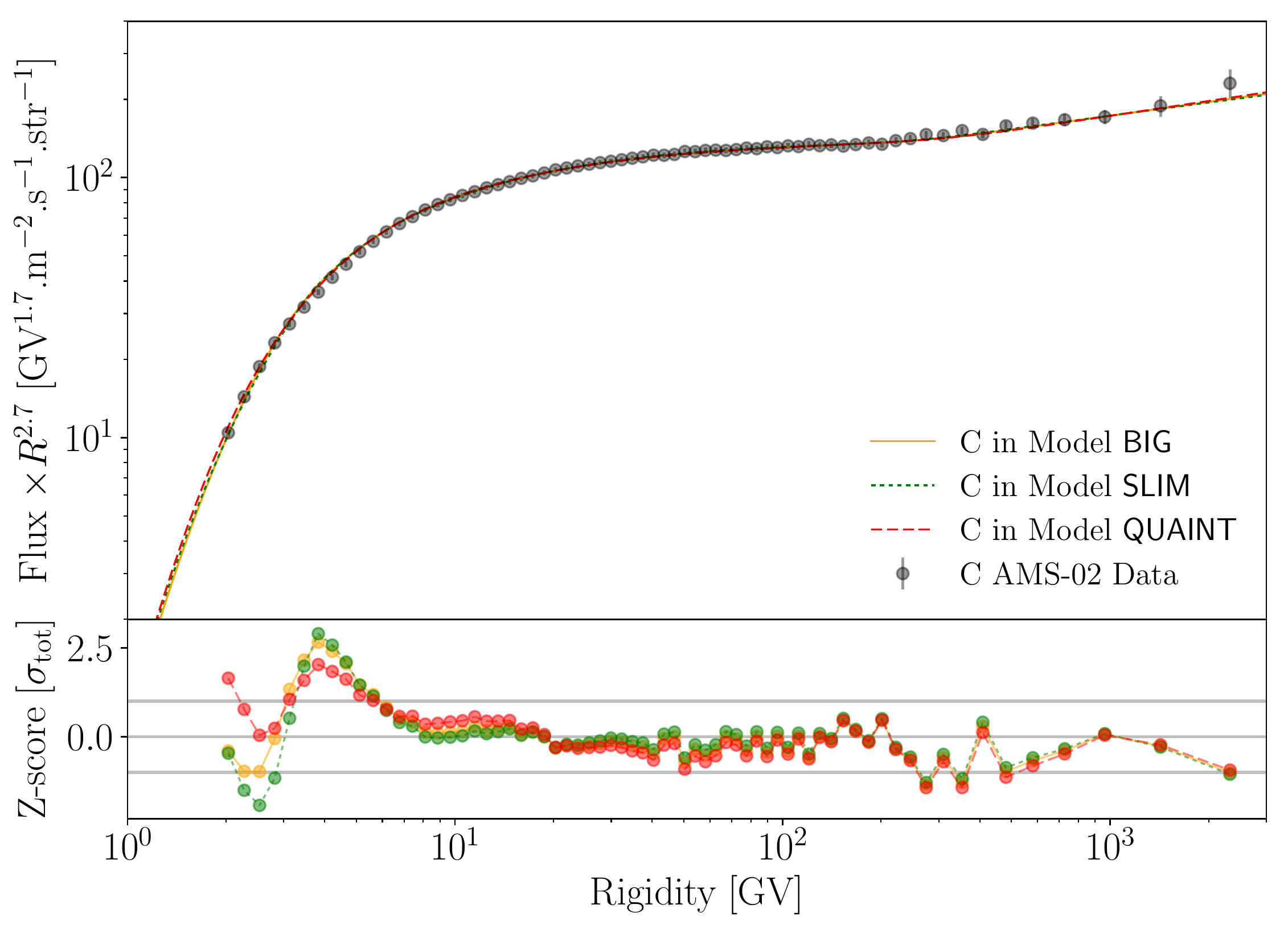}
  \includegraphics[width=\columnwidth]{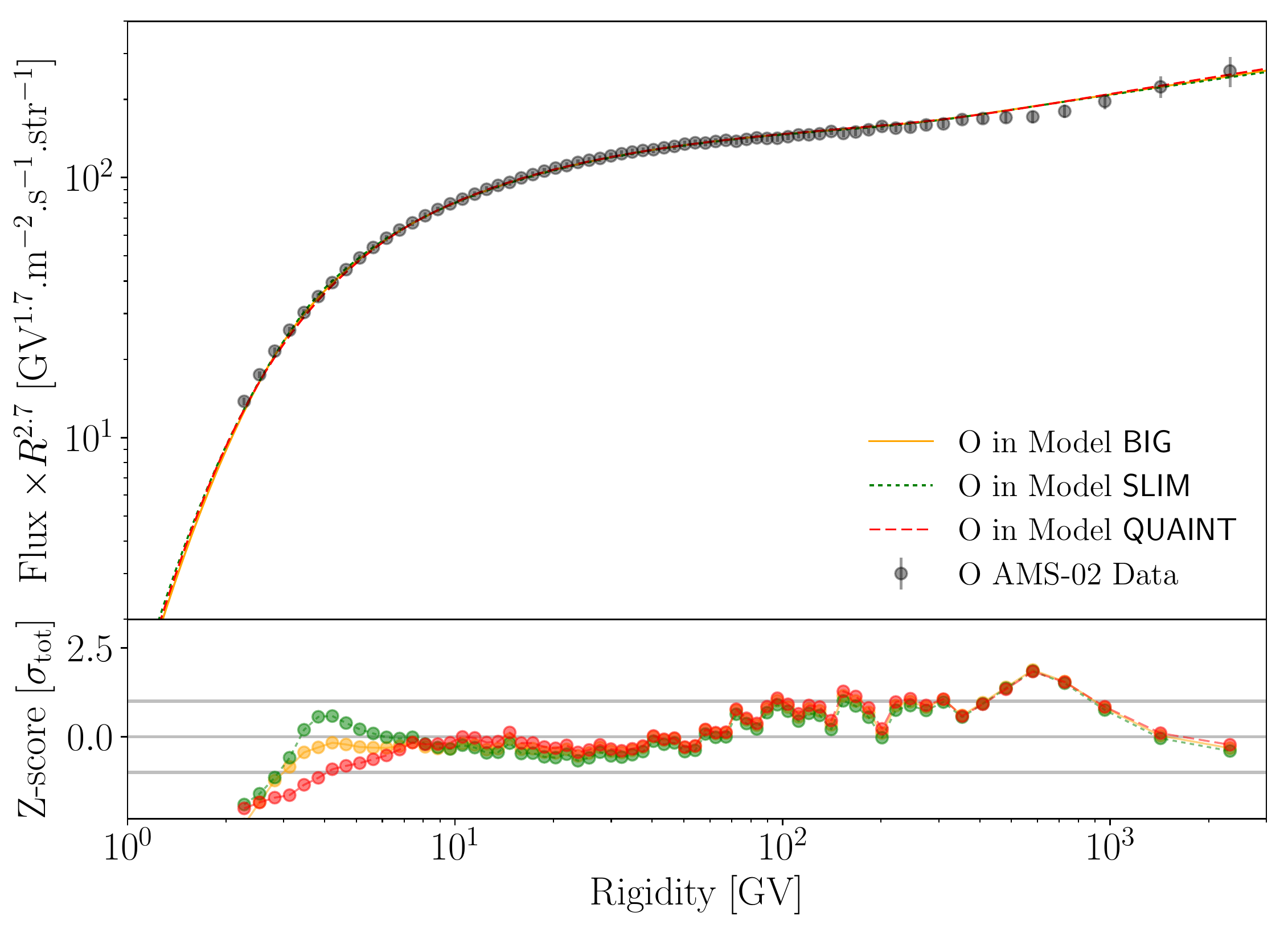}
  \caption{\footnotesize Combined best fits of C (top) and O (bottom) fluxes in the three benchmark
    models \BIG, \SLIM, and \QUAINT\ defined by B/C fits (see \citesec{sec:results}).
    A panel below each plot reports the $Z$-score,
    corresponding to the residuals normalized by the {\em total} errors $\sigma_{\rm tot}$.}
  \label{fig:best_prim_modelB}
\end{figure}

For the fits of the C and O fluxes, a simpler yet sufficient approximation is to assume uncorrelated
\textit{total} errors $\sigma_{\rm tot}$, i.e. statistical and systematic errors summed in quadrature:
on the one hand, only statistical uncertainties dominate around the high-energy break position, so
that this is a reasonable approximation. On the other hand, this fit only enters the B/C analysis
via the treatment of the high-rigidity parameters as nuisance. 

In \citefig{fig:best_prim_modelB} we report the fits of the source and high-rigidity break
parameters to the C (top panel) and O (bottom panel) fluxes for our three benchmarks, the \BIG,
\SLIM, and \QUAINT\ models. It is clear even by visual inspection that the fits with a simple,
common power-law index $\alpha$ are excellent: The fits fall within one $\sigma_{\rm tot}$ and never
beyond two $\sigma_{\rm tot}$'s from all intermediate and high-rigidity points, showing that our
consistency check is successful.
Some minor discrepancy at low rigidity is noticeable, but not worrisome for our purposes. In fact, should one aim at describing C and O primary fluxes in detail down to low rigidities, a more accurate fitting procedure treating cross-section parameters as nuisance and accounting for bin-to-bin correlations of the systematic errors (as done for B/C) would certainly reduce these minor disagreements. This is beyond our goals here, but will be of interest for future more global analyses. 
%
%
%
%
\section{Results}
\label{sec:results}
In this section, we illustrate our results for the (low- and intermediate-rigidity) propagation
parameters and discuss their implications. Initialization files used for the analysis, along with
the resulting best-fit values and covariance matrix of best-fit parameters will be provided with
the forthcoming new release \usine.
\subsection{Best-fit values and $1\sigma$ uncertainties}
The best-fit values and errors on the three model parameters (\BIG, \SLIM, and \QUAINT) are reported
in \citetab{keyres}. In the first block, we report the diffusion parameters $\delta$ and $K_{10}$
common to all models, which control the \textit{intermediate-rigidity} regime. We then report the
\textit{low-rigidity} parameters, which are different (both in nature and number) between \QUAINT,
on one side, and, \BIG\ and \SLIM, on the other. The high-rigidity break parameters, fixed following
the nuisance procedure, are reported at the bottom of the Table. The range over which we scan for
them will be discussed in \citeapp{app:consistency}, since their determination is affected by the
inclusion of external data (in our case, C and O absolute fluxes). 

In all these fits, nuisance parameters vary within reasonable pre-assigned intervals. The solar
modulation parameter  $\phi_{\rm F}$ attains a value of  731, 734 and 725  MV in the best-fit model
\BIG, \SLIM, and \QUAINT, respectively. Concerning the nuisance of the spallation cross section
$\rm ^{12}C+H \rightarrow {}^{11}B $, its best \textit{normalization} is found to be 12\%, 13\% and 11\%
above the reference GP17 value in the best-fit model \BIG, \SLIM, and \QUAINT, respectively. {The preferred \textit{slope} encoding the low energy shape is of 0.12, 0 and 0.16, for the same models. The induced spectral distorsions in \BIG\ and \QUAINT\, correspond to a slight decrease of the cross section at low energy.}

\begin{table}
  \begin{center}
    \begin{tabular}{l c c c}
 \hline\hline
  Parameters & \BIG & \SLIM & \QUAINT \\
 \hline\hline
 $\chi^2/{\rm dof}$                            &       $61.7/61\!=\!1.01$                          &     $61.8/63\!=\!0.98$    & $62.1/62\!=\!1.00$     \\[1mm]
\hline
 \multicolumn{4}{c}{Intermediate-rigidity parameters}\\
 {$K_{10}$}\,[kpc$^2$\,Myr$^{-1}$]                  &           $0.30_{-0.04}^{+0.03}$                  &   $0.28_{-0.02}^{+0.02}$       &   $0.33^{+0.03}_{-0.06} $  \\[1mm]
 $\delta$                                           &         $0.48_{-0.03}^{+0.04}$         &   $0.51_{-0.02}^{+0.02}$       &   $0.45_{-0.02}^{+0.05} $  \\
\hline
 \multicolumn{4}{c}{Low-rigidity parameters}\\
 $V_{\rm c}$   [km\,s$^{-1}$]                     &                $0^{+7.4}$                               &           N/A                   &   $0.0^{+8}$         \\
 $V_{\rm A}$   [km\,s$^{-1}$]                     &                   $67^{+24}_{-67}$                            &           N/A                   &   $101_{-15}^{+14}$           \\
 $\eta$                                  &              1 (fixed)                            &        1 (fixed)               &   $-0.09_{-0.57}^{+0.35}$    \\[1mm]
  $\delta_{\rm l}$                            &          $-0.69^{+0.61}_{-1.26}$              &      $-0.87^{+0.33}_{-0.31}$    &   N/A            \\
 $R_{\rm l} $ [GV]                           &             $3.4_{-0.9}^{+1.1}$               &     $4.4_{-0.2}^{+0.2}$        &   N/A      \\
\hline
\hline
 \multicolumn{4}{c}{High-rigidity break parameters}\\
 \multicolumn{4}{c}{(nuisance parameters)}\\
     $\Delta_{\rm h}$         &      $0.18$     & $0.19$       & $0.17$          \\
    $R_{\rm h} $ [GV]       &      $247$         & $237$          & $270 $            \\
   $s_{\rm h}$           &      $0.04$   & $0.04$  &  $ 0.04$             \\ 
 \hline
 \hline
    \end{tabular}
    \caption{Best-fit parameter values and uncertainties for the three
      benchmark models \BIG, \SLIM, and \QUAINT\ and corresponding $\chi^2/{\rm dof}$. The high-rigidity break parameters are nuisance parameters in the fit (see also text and \citeapp{app:consistency}), and their preferred post-fit values are also quoted for the sake of completeness. Errors in italic are those that reach the allowed boundaries.
      }
    \label{keyres}
  \end{center}
\end{table}

\begin{figure}[!th]
  \centering
  \includegraphics[width=\columnwidth]{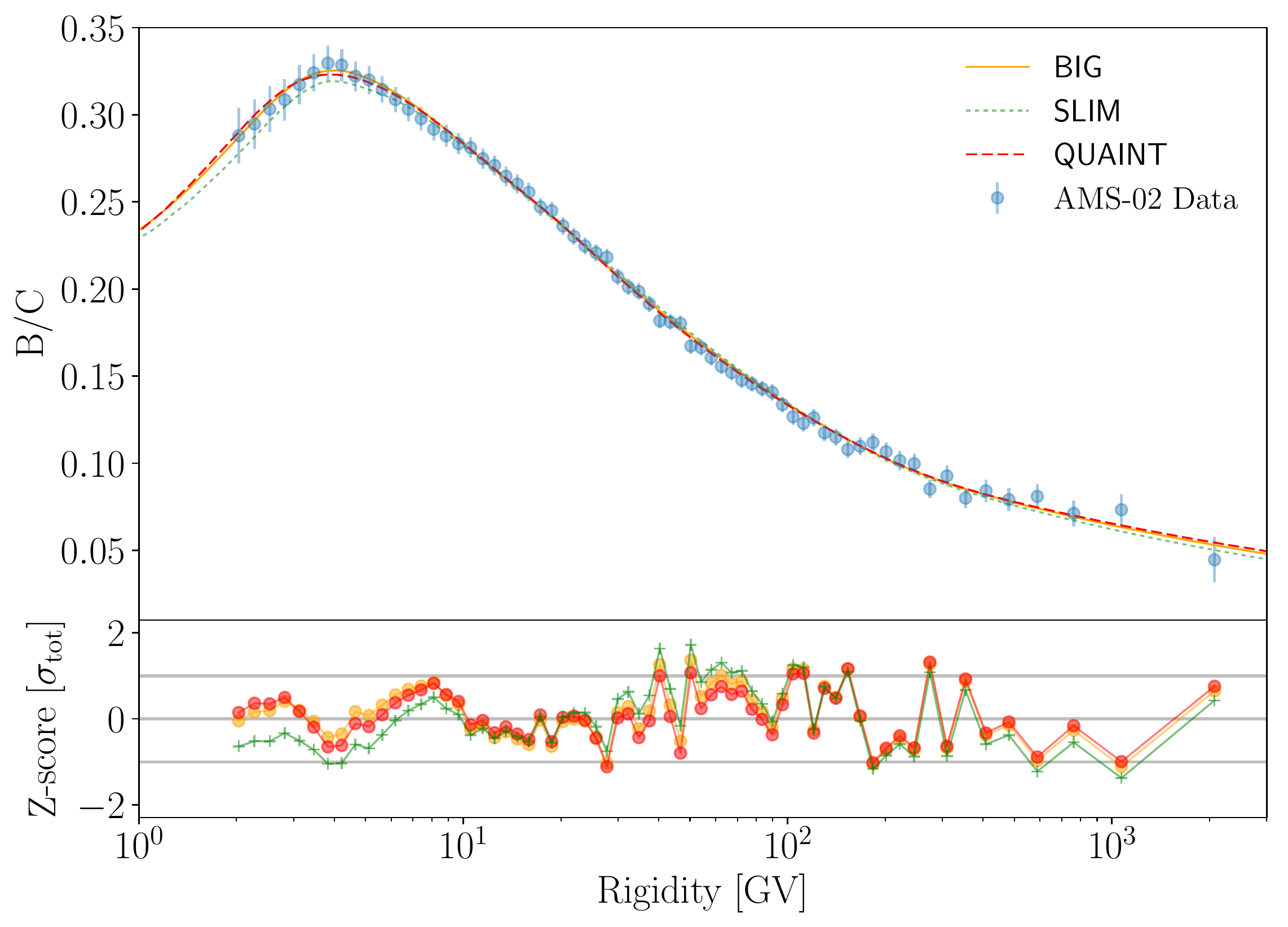}
  \caption{\footnotesize Best fit B/C curve for models \BIG, \SLIM, and \QUAINT. Results for the
    best fit parameter values are given in \citetab{keyres}. The bottom panel shows the $Z$-score.}
  \label{fig:best_fit_model}
\end{figure}

Our best-fit curves are reported in \citefig{fig:best_fit_model} for the three models. Note that
all models lead to analogous curves and fit quality, only differing in the fine features of the
spectral shape at low rigidity. The inset displays the $Z$-score, i.e. the residuals normalized to the \textit{total} errors $\sigma_{\rm tot}$. Note that this has only a qualitative purpose, since technically the $\chi^2$ is computed accounting for correlations in the systematics of B/C data, a major novelty of this analysis. The similar fit quality of the \BIG\ and \SLIM\ models indicates
that the additional free parameters present in the former are actually unnecessary to describe the
data: If the fit allows for a low-rigidity break, there is but a minor and currently unnecessary
role played by $V_{\rm c}$ and $V_{\rm A}$. We note a tiny and statistically insignificant preference
for model \SLIM\ (and a fortiori \BIG) with respect to \QUAINT, which is only worth noticing since
\QUAINT\ has one free parameter more than in \SLIM.  In fact, we stress that if we had fixed
$\eta=1$ in the \QUAINT\ model, its fit quality would have degraded, and it would have been
rejected at $>2\,\sigma$ with respect to the \BIG\ and \SLIM\ models. 
{Finally, we note that, compare to \SLIM, the benchmark \BIG\ and \QUAINT\ have respectively a weaker and no break at low rigidity, althought the latter is partly mimicked by the spectral distorsions of the cross section in nuisance. This tends to provide additional support to the possible presence of a low-rigidity break in the diffusion coefficient.}

\begin{figure}[!th]
  \centering
  \includegraphics[width=\columnwidth]{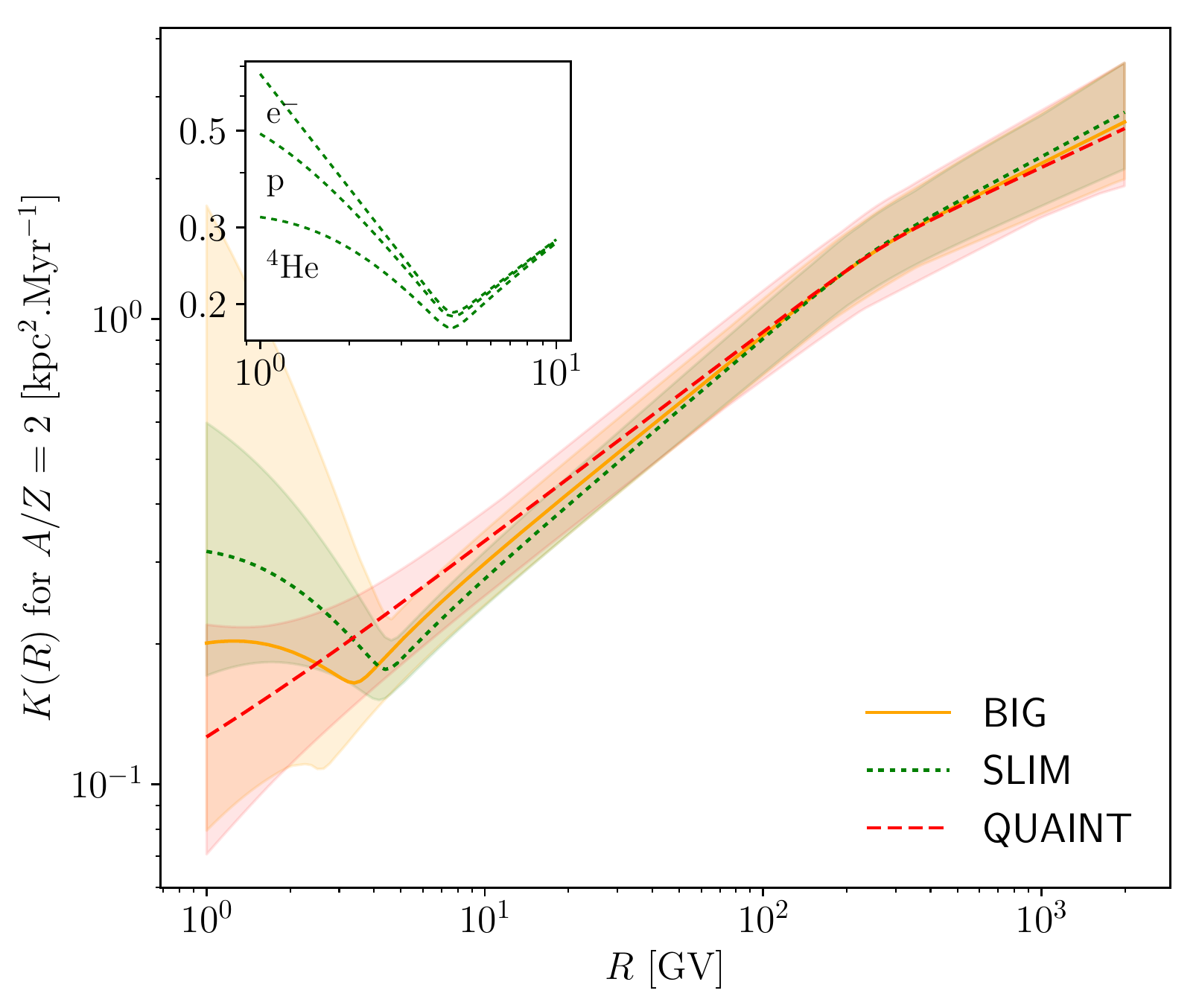}
  \caption{\footnotesize {The diffusion coefficients corresponding to the three fits
      reported in \citetab{keyres} and~\citefig{fig:best_fit_model}, with their associated 1$\sigma$ uncertainty bands. We stress that the scaling at
      low rigidity depends on where the onset of the non-relativistic regime is located; here the
      curves are traced for a mass/charge ratio $A/Z=2$. For illustration, the inset shows the low rigidity regime for various species with different mass/charge values in the \SLIM\ model.}}
  \label{fig:diffusion}
\end{figure}

Also, it is important to notice that the parameters common to the three models are found
with values compatible within $\sim 1\sigma$. This suggests that the diffusive properties at
intermediate rigidities are constrained rather robustly by the data
{(see~\citefig{fig:diffusion} for an illustration of this)},
independently of the specific scenario within which the low-rigidity behavior is interpreted and
fitted. This conclusion is rather encouraging when one considers interpretations of the
high-rigidity spectral break.

The \QUAINT\ model comes out with a few apparently surprising features: at face value, the best
fit for $V_{\rm A}$ is rather large, perhaps even more so in the light of the value found for
$V_{\rm c}$ which is compatible with zero. A too large value for $V_{\rm A}$ would lead to the
surprising conclusion that the power in diffusive reacceleration is comparable to the total CR
luminosity. This would for instance imply that the bulk of CRs energy rather originates from
ISM turbulence than from shocks in SNRs, as customarily assumed (see e.g.\citep{DruryEtAl2017}).
This apparent conundrum is alleviated once accounting for the effective nature of our
parameterization of the reacceleration term, which for technical reasons is artificially pinched to
the thin disk, rather than being present in the whole propagation halo. The actual power in the
turbulence in the whole Galaxy is thus reduced by the ratio $h/z_{\rm A}\simeq {\cal O}(h/L)$
(see \citesec{sssec:kdiff}), hence resulting roughly consistent with expectations, and also more
in line with the allowed range for $V_{\rm c}$. Another perhaps surprising outcome is the value of
$\eta$, whose best fit is {\it negative}, and anyway rather away from typically considered values
$\simeq 1$.
This conclusion is also qualitatively valid in models \BIG\ and \SLIM: there is a slight preference
for the diffusion coefficient below about 4.5 GV to {\it increase with lower rigidity}. Note that,
within the allowed range for $\delta_l$, extrapolation in the non-relativistic regime could lead to
nonphysical results, as soon as $\delta_l+1=\delta-\Delta_{\rm l}+1<0$.
Be that as it may, neither this caveat nor the previous one for $\eta$ in the \QUAINT\ model
should be overstated, since they arise at best at the $1\sigma$ level. All model fits are consistent
with a perhaps more physically acceptable flat behavior, or a rigidity-independent diffusion
coefficient at low-$R$. Furthermore, it is worth noting how the \BIG\ model is closer to the
relatively unproblematic regime $\delta_{\rm l}>-1$ than the \SLIM\ one. The introduction of
some reacceleration and convection (both physically expected) tends to yield more reasonable values
for the low-rigidity slope.
\subsection{Possible interpretation and microphysics}
In any case, the most obvious interpretation of these results is that there are less and less
waves onto which CRs can scatter at low rigidity. One possible reason is that turbulence dissipation
effects lead to a parallel diffusion coefficient which decreases with increasing rigidity, with
turning point at $\sim 3\;$GV  for plausible choices for the parameters~\cite{ShalchiEtAl2010}. 
Another possibility arises in models where the CRs scatter onto self-generated turbulence below some
rigidity (see e.g. \citep{BlasiEtAl2012a}). The energy density (and the pressure) carried by CRs
peaks at the few GV scale; above this rigidity, the induced diffusion coefficient increases with
rigidity as customarily assumed, because of the relatively steep CR power-law spectrum. Below this
rigidity, however, the lower the rigidity (or Larmor radius), the smaller the turbulence with
respect to extrapolations, simply because there are less and less CRs that can generate it by
streaming instability due to their spectral inflection. The order of magnitude of the break in the
low-energy CR spectrum seems to be in the right ballpark, but these qualitative arguments deserve a
more detailed investigation, which we postpone to future work.
\subsection{Robustness of low-, intermediate-, and high-rigidity parameters}
A very encouraging finding is that, within uncertainties, the diffusive properties at intermediate
rigidities do not depend on the specific scenario considered at low-rigidity. The value found for
$\delta$ appears closer to a Kraichnan turbulence spectrum ($\delta\simeq 0.5$) than to a Kolmogorov
one ($\delta\simeq 1/3$), although this conclusion should not be overstated since the model involves
an {\it effective} isotropic diffusion coefficient. An indirect implication of this robustness is
to increase the credibility in any deviation found at high rigidity, of course.

Concerning the low-rigidity regime, however, there are several important caveats, which suggest some
prudence to avoid over-interpreting the values found. First of all, while there is a clear
indication for a different regime of propagation at low rigidity, the ``hardest'' parameters to
interpret ($\eta$ and $\delta_{\rm l}$) are actually heavily influenced by the one or two
lowest-rigidity points. This is illustrated in more detail in \citeapp{app:low_rig}, where one
can compare the behavior of $R_{\rm l}$ vs. $R_{\rm min}$ with respect to $\delta_{\rm l}$ vs.
$R_{\rm min}$, $R_{\rm min}$ being the rigidity above which the fit is performed. There is simply not enough of a baseline at low rigidity in the AMS-02 data to
unambiguously measure the slope in this range. Another point to keep in mind is that the
low-rigidity range is quite influenced by the uncertainties in the nuclear cross sections and
the treatment of solar modulation. Indeed, including the nuisance parameters for the production cross section increases the $1\sigma$ uncertainties on $\eta$ (\QUAINT) by 50\%, and on $R_l$ and $\delta_l$ (\BIG) by 90\%. In our fits, including solar modulation is a second order effect, since it increases the low-energy parameters uncertainties by order 5\%.

The only model-independent conclusion that we can safely make on the low-rigidity range is that
multiple models can account for the observations, with rather different physical interpretations
possible. So, statements such as ``the reacceleration/convection velocity determined from the
B/C data is \ldots'' should be taken with a grain of salt, since they appear {\it very} model
dependent, if compared, for instance, with the determination of $\delta$. The fitted values should
{\it only} be used as references in the same model used to fit them, and extrapolations at lower
rigidities (below the range covered by the data) are not guaranteed to be physical.

\section{Summary, conclusions, and perspectives}
\label{sec:conclusion}
This article has set the stage for the propagation scenarios that we want to test, challenge and
refine with further AMS-02 data, defining benchmark models and ranges of parameters.  We have
validated the first step of this program with a statistically more sound analysis of the AMS-02 B/C
data, going beyond state-of-the-art in the modern literature, and checking different theoretical
frameworks differing in the treatment of transport at low rigidities with a major (model \QUAINT)
or a negligible (model \SLIM) role played by reacceleration. Both models are limiting cases of a
more general model (\BIG). We have made sure that issues like numerical stability, the effects of
cross sections uncertainties, the bin-to-bin correlation of systematic errors are handled
sufficiently well not to bias significantly the conclusions. 

For the time being, either model can describe with comparable performances the low-rigidity regime, with a statistically insignificant preference for model \SLIM. The parameters describing intermediate rigidities are consistently determined in either case. At low rigidity, degeneracies with nuisance parameters impact both the best fit and uncertainties, in particular the ones controlling the energy shape of cross sections and solar modulation. This means that qualitatively different models offer almost equally good description of the data, so that inferring the physics of the propagation at low rigidity is challenging, and we must content ourselves with one or another ``effective'' description. This lesson on the shaky discrimination power among models with mild differences at low rigidities is likely to apply more generally, even to alternative models not tested here, because it partially relies on the effect of the nuisance parameters. Obviously, any discrimination between the two sets of models must be based on complementary data or arguments, such as the (astro)physical plausibility of the parameters found, an issue which we also briefly discussed. However, finding a break
in the diffusion coefficient at low rigidity should not come so much as a surprise, since this
feature, possibly related to some damping in the turbulence spectrum and the subsequent increase of
the CR mean free path, is expected from theoretical grounds \cite{YanEtAl2004,PtuskinEtAl2006a,ShalchiEtAl2010,EvoliEtAl2014}. In this respect, a careful study of low energy data complimentary to AMS-02 ones (e.g. from ACE-CRIS~\cite{LaveEtAl2013} and Voyager I~\cite{CummingsEtAl2016,WebberEtAl2017}), together with a more realistic account of the systematic error correlations (based on further information provided by the experimental collaborations), could certainly help in drawing more robust conclusions on the properties and the nature of this break.

Besides extracting reference propagation parameters and uncertainty ranges from B/C data, which are intended for references for further studies, we have also performed a first test of the consistency of the obtained results with simple CR source spectra (power laws). We also confirmed and strengthened our conclusions in~\cite{GenoliniEtAl2017}, that the high-rigidity data can be consistently interpreted as a consequence of a break in the diffusive coefficient, in agreement with AMS-02 high-energy primaries spectra. Indeed, we show that this preference does persist in a generalized analysis extending to the whole rigidity range, and for alternative propagation setups, notably with/without reacceleration.

Following this study, the most pressing issue is of course to test the reference models provided here  against other secondary data (e.g., Li, Be, pbar, positrons). In particular, our forthcoming publication will focus on the antiproton channel \cite{BoudaudEtAl2019}. There has been a recent interest in the possibility that these data hide a signal of dark matter annihilation, see e.g.~\cite{CuocoEtAl2019,CholisEtAl2019}, and it is interesting and important to re-examine those claims within our analysis framework. Finally, it is also known that putative dark matter signals are sensitive to the diffusive halo size, hence an important and motivated follow-up project analysis will involve other secondaries, including  isotopes such as the radioactive species (e.g. ${}^{10}$Be). To that purpose, we provide the reader in \citeapp{app:Ldiff} (1D models) and in \citeapp{app:dict} (2D models) with the scaling relations that allow to extrapolate our benchmark models (derived assuming $L=10$~kpc) to a range of $L$ between 4 and 18 kpc.

{\it Note added:} As we were completing this study, we became aware of \cite{VittinoEtAl2019}, where the authors find
 support for the presence of multiple breaks in the diffusion coefficient, based on CR electron and positron data.
This result is complementary to (and consistent with) ours, while interestingly based on independent datasets concerning other CR species.

\begin{acknowledgments}
  This work has been supported by the ``Investissements d'avenir, Labex ENIGMASS",  by the French
  ANR, Project DMAstro-LHC, ANR-12-BS05-0006, and by Univ. de Savoie, AAP ``DISE''. It has also
  benefited from the support of the ANR project GaDaMa (ANR-18-CE31-0006), the OCEVU Labex
  (ANR-11-LABX-0060), the CNRS IN2P3-Theory/INSU-PNHE-PNCG project ``Galactic Dark Matter'', and
  European Union's Horizon 2020 research and innovation program under the Marie Sk\l{}odowska-Curie
  grant agreements N$^\circ$ 690575 and No 674896. The work of Y.G. is supported by the IISN, the FNRS-FRS and a ULB ARC. The work of M.B. is supported by the European Research Council ({ERC}) under the EU Seventh Framework Program (FP7/2007-2013) / {ERC} Starting Grant (Agreement No 278234 -- {"NewDark"} project).
\end{acknowledgments}

\appendix
\section{On the high-rigidity break from C, O and the fitting procedure}
\label{app:consistency}

The fitting procedure described in \citesec{sec:fit_bc} makes use of the C and O fluxes:
\begin{enumerate}
\item As a sanity check for the actual diffusion parameters inferred (see \citefig{fig:best_prim_modelB}).
\item To determine $\alpha$, the common spectral index for all nuclei, although its value is irrelevant for the B/C calculation (and for the transport parameter determination).
\item Above all, for cornering a plausible window for the nuisance of the the high-energy break parameters. 
\end{enumerate}
Below, we provide some consistency checks as well as some comments on these ancillary results. \\ 
\subsection{Consistency check}
In \citetab{keyresH}, we report the results of the reference fits of C, O fluxes corresponding to the B/C model fits discussed in \citesec{sec:def_bench}. 
\begin{table}[!htb]
  \begin{center}
    \begin{tabular}{|c || c | c |c|  }
      \hline\hline
      Parameters & {\sf BIG} & {\sf SLIM} &  {\sf QUAINT} \\
    \hline\hline
   $\chi^2/{\rm dof}$       &       $75.7/129= 0.59$        & $73.2/129=0.57$          &        $80.3/129= 0.62$                           \\
    $\alpha$                &      2.35
    &  2.33    
    & 2.36     
    \\
     $\Delta_{\rm h}$         &      $0.18_{-0.05}^{+0.13}$     & $0.18^{+0.11}_{-0.04}$       & $0.18_{-0.01}^{+0.18}$          \\
    $R_{\rm h} $ [GV]       &      $244_{-52}^{+198}$      & $236_{-51}^{+152}$          & $282_{-89}^{\textit{+349}} $            \\
   $s_{\rm h}$        &      $0.04_{\it -0.04}^{+0.11}$   & $0.03_{\textit{-0.03}}^{+0.09}$  &  ${ 0.04}^{+0.15}_{\textit{-0.04}}$             \\ 
    \hline
    \end{tabular}
    \caption{The $\chi^2/{\rm dof}$, the best-fit value for $\alpha$ as well as error range (used as nuisance parameters and ranges in the  B/C analysis) for the \textit{high-rigidity} parameters, coming from
     the combined fit to absolute C, O fluxes in the iterative procedure described in \citesec{sec:fit_bc}. Values in italics means that the fit reached the border of the interval. }\label{keyresH}
  \end{center}
\end{table}

All propagation models inferred from B/C appear to provide excellent fits to the C, O fluxes as
well. 
The $\chi^2$ cannot be used at face value as a quantitative estimator of the quality of the fit, since {\it total} errors have been used in the C, O fits: Hence, we likely underestimate the contribution to the $\chi^2$, notably those of the intermediate- and low-rigidity data mostly influenced by systematic errors and their correlations (see also the companion paper~\citep{DeromeEtAl2019}). Nonetheless, a relative preference seems to emerge for the {\sf BIG} and {\sf SLIM} models, compared to the {\sf QUAINT} model, which is interesting as the same trend is also present from the more rigorous B/C analysis.

Concerning $\alpha$, the values found are intriguingly similar to the ones found in the fit of the He flux, which is performed in \citep{BoudaudEtAl2019}, another reassuring consistency test of our procedure. We are thus consistent with the current universality of the spectra of nuclei (while the proton flux seems to be somewhat steeper). This is an interesting observable to keep an eye on in the future, of course. Note that the fit yields a nominal error on the parameter $\alpha$ at the sub-percent level, since $\delta$ is kept fixed in the iteration. Realistic uncertainties on $\alpha$ are however comparable to the ones of $\delta$ reported in \citetab{keyres}.

In all cases, the indication for a high-rigidity break $\Delta_{\rm h}$ is rather significant ($\gtrsim 4\,\sigma$), again consistently with AMS-02 results, but here referring to the underlying diffusive coefficient (i.e. a break in the model space, not in the flux spectral index). Also, the values found are consistent within the errors with those found from $p$, He analyses (e.g.~\citep{GenoliniEtAl2017}), although a bit higher, i.e. indicating a slightly more pronounced break. It will be interesting to follow-up on this in the light of further analyses of both light and intermediate/heavy nuclei, to see if the situation will relax towards a more common value or point to some discrepant hardening. \\

Let us briefly develop further on the significance of the high-rigidity break in the light of B/C data only. 

\begin{table}[!htb]
\begin{center}
\begin{tabular}{|c | c | c |c|}
\hline\hline
Parameters & {\sf BIG} & {\sf SLIM} &  {\sf QUAINT} \\
\hline\hline
$\delta$                    &  $0.55_{-0.04}^{+0.20}$      &   $0.55 ^{+0.09}_{-0.03}$        &   ${\it 0.9}_{-0.23}$         \\
$K_{10}$   [kpc$^2$/Myr]    &  $0.26_{-0.2}^{+0.05}$       &   $0.26 _{-0.01}^{+0.07}$    &   $0.10^{+0.07}_{-0.01}$  \\
$V_{\rm A}$   [km/s]       &  $0^{+64}$                    &           NA        &   $71_{-7}^{+20}$                 \\
$V_{\rm c}$   [km/s]       &  $0^{+16}$                     &           NA       &   $19_{-5}^{+3}$                \\
$\eta_t$                    &  1 (fixed)                           &        1 (fixed)  &   $-0.30_{-0.75}^{+0.54}$      \\

$\delta_{\rm l}$              &  $-0.84^{+0.32}_{-0.36}$              &      $-0.87^{+0.35}_{-0.33}$ &   NA                      \\
$R_{\rm l} $ [GV]               &  $4.4_{-2.1}^{+0.46}$               &     $4.4^{+0.2}_{-0.2}$  &   NA                      \\

$\Delta_{\rm h}$             &  $0.27_{-0.12}^{+0.22}$              & $0.27_{{\it -0.12}}^{+0.21}$  & $0.56_{-0.24}^{+0.09}$          \\
$R_{\rm h} $ [GV]               &  $158_{{\it -58}}^{+235}$            & $159^{+240}_{{\it -59}}$ & $ {\it 100}^{+96}$             \\
$s_{\rm h}$                     &  $0.10_{{\it -0.10}}^{{\it +0.20}}$  & $0.11_{-0.1}^{{\it +0.19}}$ &  $0.26_{{\it-0.26}}^{\it +0.04}$                  \\
\hline
$\chi^2/{\rm dof}$               & $58.6/58=1.01$                       &     $58.7/60=0.98$  & $59.7/59=1.01$               \\
\hline
\end{tabular}
\caption{{\bf }Best fit parameters for models  {\sf BIG}, {\sf SLIM}, and  {\sf QUAINT}, if fitting the high-energy break of the diffusion coefficient as well on the B/C data only. Values in italics means that the fit reached the border of the interval.}
\end{center}
\label{tab:results_break_free}
\end{table}
If we were to use solely B/C data to fit {\it also} the high-rigidity parameters (i.e. {\it without} relying on the C, O flux data), we would obtain the results listed in Table~IV. 
 The low and intermediate rigidity propagation parameters are consistent with our reference one (see \citetab{keyres}), with larger error bars, as expected since we now are determining more parameters from a more restricted set of data. Similar considerations apply to high-rigidity parameters, compare with \citetab{keyresH}. The largest departures are seen in the  {\sf QUAINT} model, where one suffers from a partial degeneracy of the (large) $V_{\rm A}$ parameter with the others, including $\delta$. Also, in this case parameters tend to drift towards the borders of the ``plausible'' interval fixed beforehand, which puts into question how physically meaningful this model results really are. Still, in all cases there is an evidence for a high-rigidity break (at $\gtrsim 2\,\sigma$ level, naively speaking) from the B/C alone, which {\it a posteriori} is  a justification for our choice of the parameterization of the diffusion coefficient, \citeeq{eq:def_K}. 
  
\subsection{Break vs no-break}
In Tab.~V 
we report the best fit propagation parameters without high-rigidity break in the diffusion coefficient. Note how the values of $\delta$ would be biased (at the $1\div 2\,\sigma$ level),  resulting in a harder diffusion coefficient.
In the same spirit as \cite{GenoliniEtAl2017}, we compute the $\Delta\chi^2$ with respect to our results in~\citetab{keyres} (break parameters fit to C and O fluxes). 
In the  {\sf QUAINT} model, as intuitively expected, the presence of a large $V_{\rm A}$ can partially mimic the break, but not completely, and the ``no break'' case is still disfavored (at $\sim 2\,\sigma$ level). For the  {\sf BIG} and {\sf SLIM} models, which are refinements of the purely diffusive intermediate/high-rigidity model considered in~\cite{GenoliniEtAl2017}, we find $\Delta\chi^2>10$, confirming (and thus reinforcing the robustness of) the results presented in \cite{GenoliniEtAl2017}.

\begin{table}[!htb]
    \begin{center}
    \begin{tabular}{|c | c | c |c|}
    \hline\hline
Parameters & {\sf BIG} & {\sf SLIM} &  {\sf QUAINT}\\
    \hline\hline
    $\delta$                &        $0.48^{+0.02}_{-0.02}$          &   $0.48^{+0.02}_{-0.02}$  &   $0.42^{+0.03}_{-0.02}$         \\
    $K_{10}$   [kpc$^2$/Myr]&         $0.29^{+0.02}_{-0.02}$    &   $0.29^{+0.02}_{-0.02}$        &   $0.36^{+0.02}_{-0.04}$                    \\
    $V_{\rm A}$   [km/s]   &         $0^{+115}$                   &           NA     &   $113^{+7}_{-15} $                  \\
    $V_{\rm c}$   [km/s]   &          $0^{+12}$                &           NA       &   $0 ^{+4.1}$                   \\
    $\eta_t$                &          1 (fixed)                   &        1 (fixed)     &   $0.6^{+0.3}_{-0.5} $       \\
    
    $\delta_{\rm l}$          &          $-0.88_{-0.30}^{+0.31}$             &      $-0.88_{-0.30}^{+0.32}$ &   NA                     \\
    $R_{\rm l} $ [GV]         &           $4.4^{+0.23}_{-2.4}$          &     $4.4^{+0.24}_{-0.21}$  &   NA                    \\
    \hline
$\chi^2/{\rm dof}$          &       $72.8/61=1.19$               &     $72.8/63=1.16$   & $67.1/62=1.08$                \\
    \hline
    \hline
    $\Delta\chi^2$          &        11.1       &     11.0  & 5.0                  \\
    \hline
    \end{tabular}
    \caption{Best fit parameters for models  {\sf BIG}, {\sf SLIM}, and  {\sf QUAINT}, with no high-rigidity break in the diffusion coefficient.}
    \end{center}
    \label{tab:results_without_break}
    \end{table}


\section{Fit parameters dependence upon low-rigidity cutoff}
\label{app:low_rig}

In \citefig{fig:evo_param_vs_Rmin_modA}, we present the evolution of the best-fit parameters as a function of a low-rigidity cut $R_{\rm min}$ above which the fit is performed, for the three models considered. We note that the value of the parameters $\delta$ and $K_{10}$ remains essentially unchanged whatever $R_{\rm min}$. In contrast, $\delta_{\rm l}$, $R_{\rm l}$ (for \BIG~and \SLIM), $V_a$, and $\eta$ (for \QUAINT) depend crucially on the first data points, notably those below  $\sim$4 GV. This explains their denomination  of \textit{low-rigidity} parameters. For the \SLIM\ model, note how the error on $\delta_{\rm l}$ crucially depends on the first couple of AMS-02 points, and the evidence for a change of slope (a determination of  $R_{\rm l}$) is stronger than the actual value of the slope at low rigidity.
Finally, it is worth commenting on $V_{\rm A}$: this parameter is  (anti)correlating with low-energy ones (in particular $\eta$ for \QUAINT) and, to a minor extent, also with $\delta$. This is not very surprising since large values of $V_{\rm A}$ imply ``cross-talk'' among energy bins. 
\label{app:primaries}
 \begin{figure*}[!th]
  \centering
  \includegraphics[width=0.64\columnwidth]{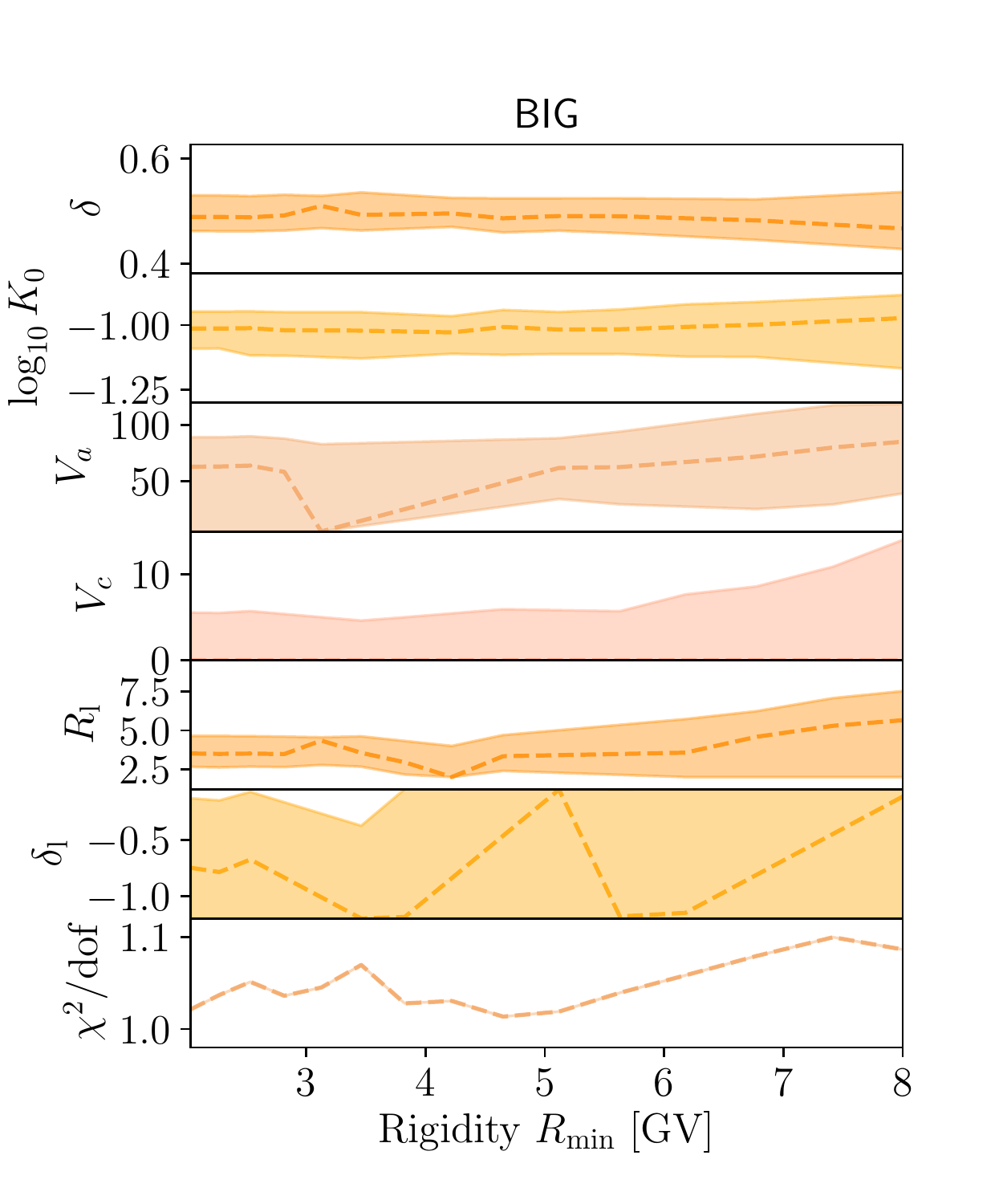}
  \includegraphics[width=0.64\columnwidth]{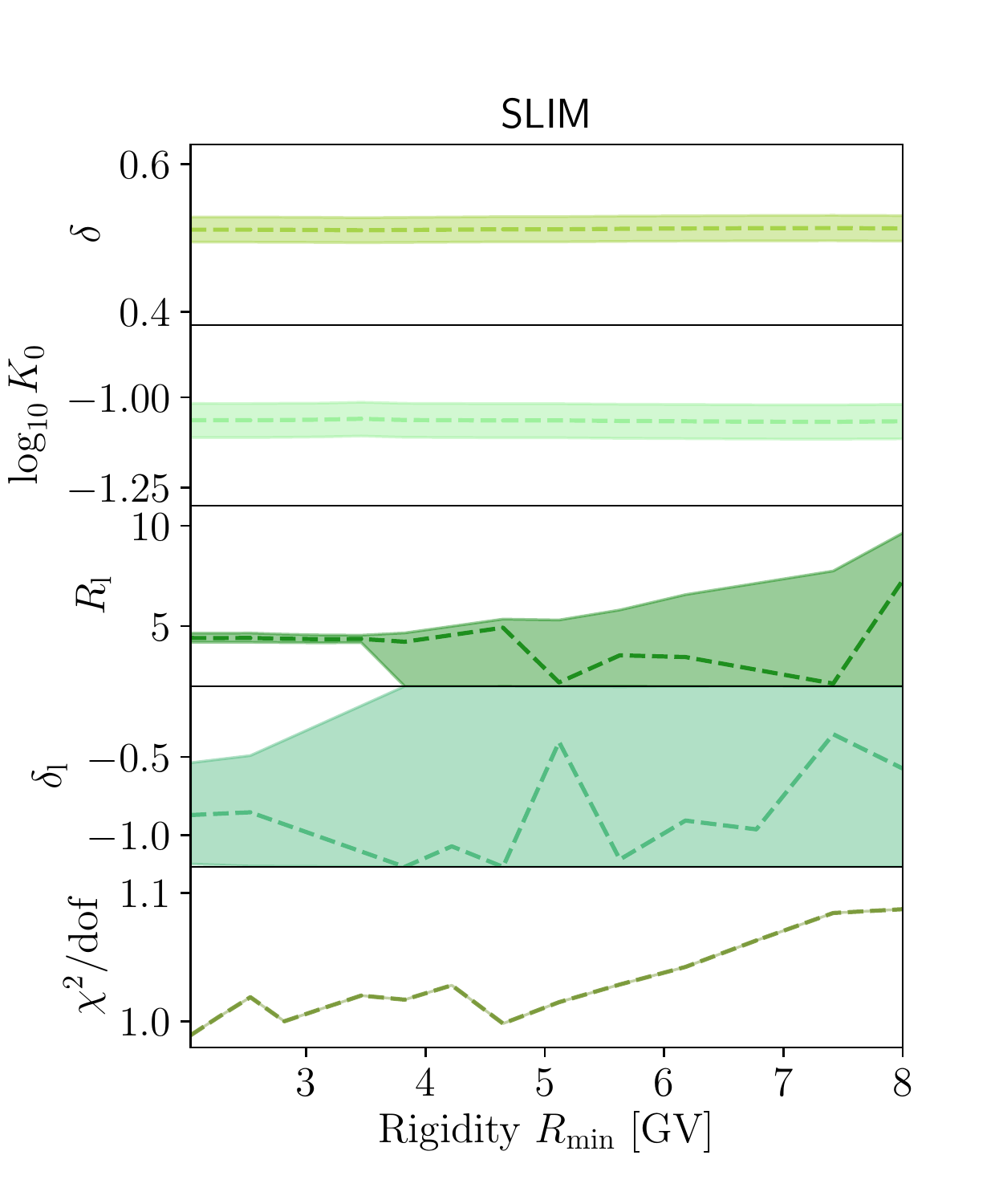}
  \includegraphics[width=0.64\columnwidth]{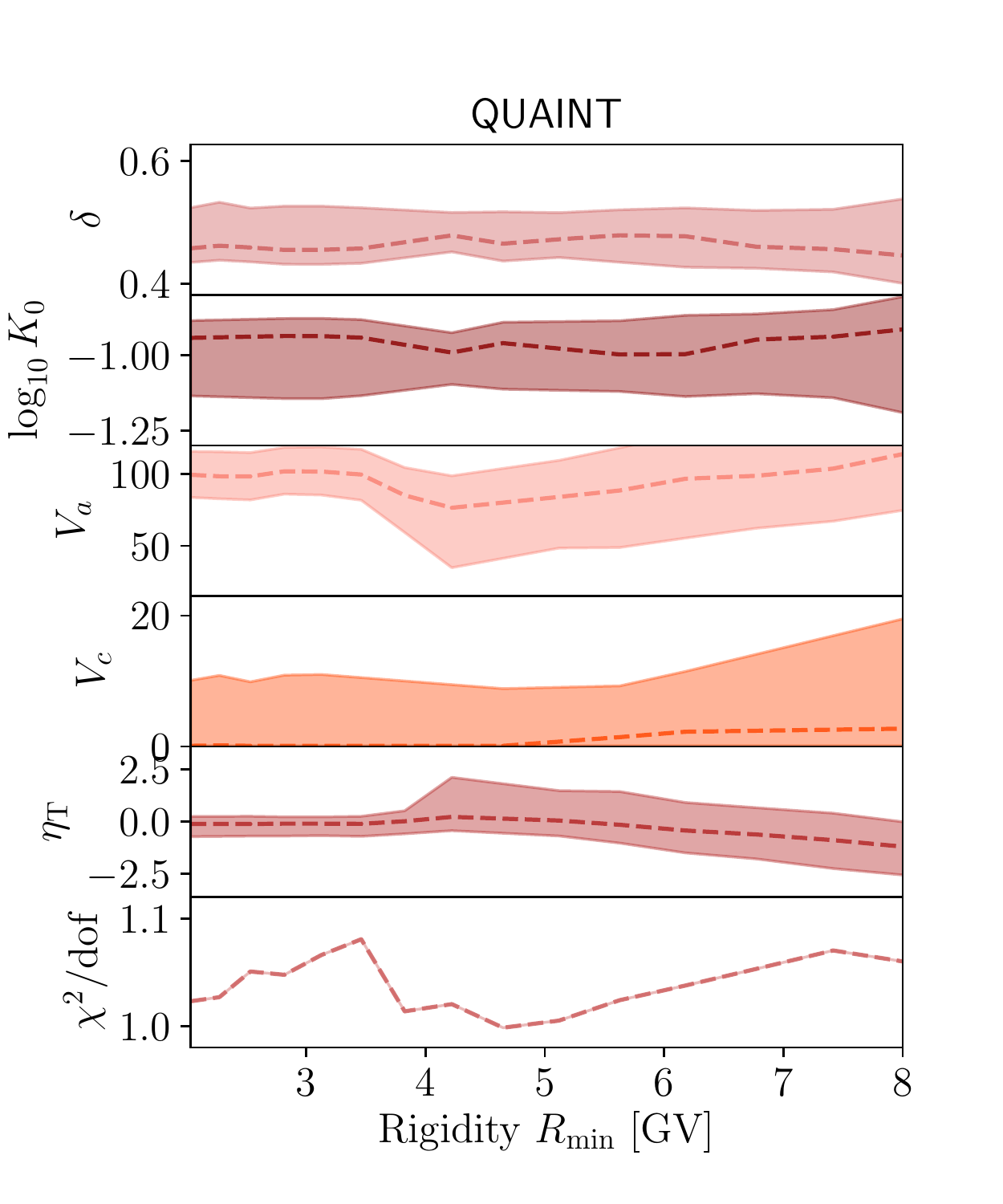}
  \caption{\footnotesize Evolution of the best fit parameter values and uncertainties as a function of the minimal rigidity $R_{\rm min}$ above which the fit is performed, for the three benchmark models \BIG, \SLIM\ and \QUAINT. Note that in this figure only we use $K_0$ (normalization of $K$ at $R=1\,$GV) instead of $K_{10}$.}\label{fig:evo_param_vs_Rmin_modA}
\end{figure*}

\section{Scaling of propagation parameters with $L$ in 1D model}
\label{app:Ldiff}

The benchmark parameters for \BIG, \SLIM, and \QUAINT\ where derived assuming the Galactic magnetic halo shapes as a 1D-slab of half-thickness $L=10$ kpc. By fitting the B/C ratio in these models, it is well known that the normalization of the diffusion coefficient $K_{10}$ and the halo thickness $L$ are degenerated so that the ratio $K_{10}/L$ is constant. We have checked that this was still the case given the higher sensitivity of AMS-02 data, and found the following scaling relations for values of $L$ within [4,18] kpc:

\besub
\label{eq:sc_relation}
\begin{eqnarray}
&&\text{\BIG:  }\quad\frac{K_{10}}{L}=0.030^{+0.003}_{-0.004}\;\rm kpc/Myr\;.\\
&&\text{\SLIM:  }\quad\frac{K_{10}}{L}=0.028^{+0.002}_{-0.002}\;\rm kpc/Myr\;,\\
&&\text{\QUAINT:  }\quad\frac{K_{10}}{L}=0.033^{+0.003}_{-0.006}\;\rm kpc/Myr\;.
\end{eqnarray}
\eesub
 
\section{Dictionary to use 1D propagation parameters in 2D models.}
\label{app:dict}

A 1D-slab geometry for the magnetic halo does not allow one to account for CRs escaping radially from the Galaxy (see, e.g. refs \cite{Taillet:2002ub,Maurin:2002uc}). This choice could be thought as an over-simplification. A more realistic 2D geometry commonly used is to consider the Galactic halo as a cylindrical box of radius 20 kpc and half-thickness $L=10$ kpc, where CR sources lie uniformly in the disk and the Earth is set at 8.5 kpc from its center (for an illustration see e.g. Fig.10 of \cite{GenoliniEtAl2015}). However, using this geometry we have found that variations of the best fit values (\citetab{keyres}) for all parameters, except $K_{10}$, are within their respective uncertainties. In fact, in this 2D (uniform disk) case, a degeneracy between $K_{10}$ and $L$ is still present, but is no longer described by the relation \citeeq{eq:sc_relation}; the escape from the radial boundaries increases with increasing $L$. Starting from \citetab{keyres}, for each benchmark, we summarize below our empirical prescription to go from 1D to 2D, only for the parameters which drift with $L$. Note that the preferred value for $V_{\rm A}$ in the \BIG~model is now zero, although the uncertainty on this parameter is quite large.

\begin{widetext}
\besub
\begin{eqnarray*}
&& \text{\BIG: }\,\frac{K^{\rm 2D}_{10}}{\tanh{(L^{1.1}/10.1)}}=0.25^{+0.04}_{-0.02}\,{\rm kpc^2/Myr}, \quad
\delta^{2D}\times \tanh{(L^{0.4}/0.77)}=0.50^{+0.02}_{-0.04}\quad\text{and} \quad \begin{cases} V_{\rm A}=0^{+80} \,{\rm km.s^{-1}} \\ R_{\rm l}=4.4_{-0.2}^{+0.2} \,{\rm GV}  \\ \delta_{\rm l}=-0.83_{-0.3}^{+0.3} \end{cases}\,,\\
&& \text{\SLIM:  }\,\frac{K^{\rm 2D}_{10}}{\tanh{(L^{1.1}/10.1)}}=0.25^{+0.02}_{-0.02}\,{\rm kpc^2/Myr}\quad\text{and} \quad \delta^{2D}\times \tanh{(L^{0.4}/0.77)}=0.50^{+0.02}_{-0.02}\,,\\
&& \text{\QUAINT:  }\,\frac{K^{\rm 2D}_{10}}{\tanh{(L^{1.1}/10.1)}}=0.30^{+0.02}_{-0.07}\,{\rm kpc^2/Myr}\quad\text{and}\quad\delta^{2D}\times \tanh{(L^{0.4}/0.77)}=0.44^{+0.06}_{-0.02}\,.
\end{eqnarray*} 
\eesub
\end{widetext}

\bibliographystyle{apsrev4-1}
\bibliography{BCprd}
\end{document}